\documentclass[5p,sort&compress,hidelinks]{article}

\usepackage{subfig}
\usepackage{hyperref}
\usepackage{listings}
\usepackage{arxiv}

\usepackage{graphicx}
\usepackage{booktabs}
\AtBeginDocument{%
  }

\usepackage{listings}

\usepackage{textcomp}
\usepackage{multirow}

\usepackage{xcolor} 
\definecolor{azure}{rgb}{0.0, 0.5, 1.0}
\definecolor{darkgreen}{rgb}{0.0, 0.5, 0.0}
\definecolor{amaranth}{rgb}{0.9, 0.17, 0.31}
\definecolor{cadetgrey}{rgb}{0.57, 0.64, 0.69}
\definecolor{aureolin}{rgb}{0.99, 0.93, 0.0}
\definecolor{amber}{rgb}{1.0, 0.49, 0.0}

\newcommand{\orcid}[1]{\href{https://orcid.org/#1}{\includegraphics[height=10pt]{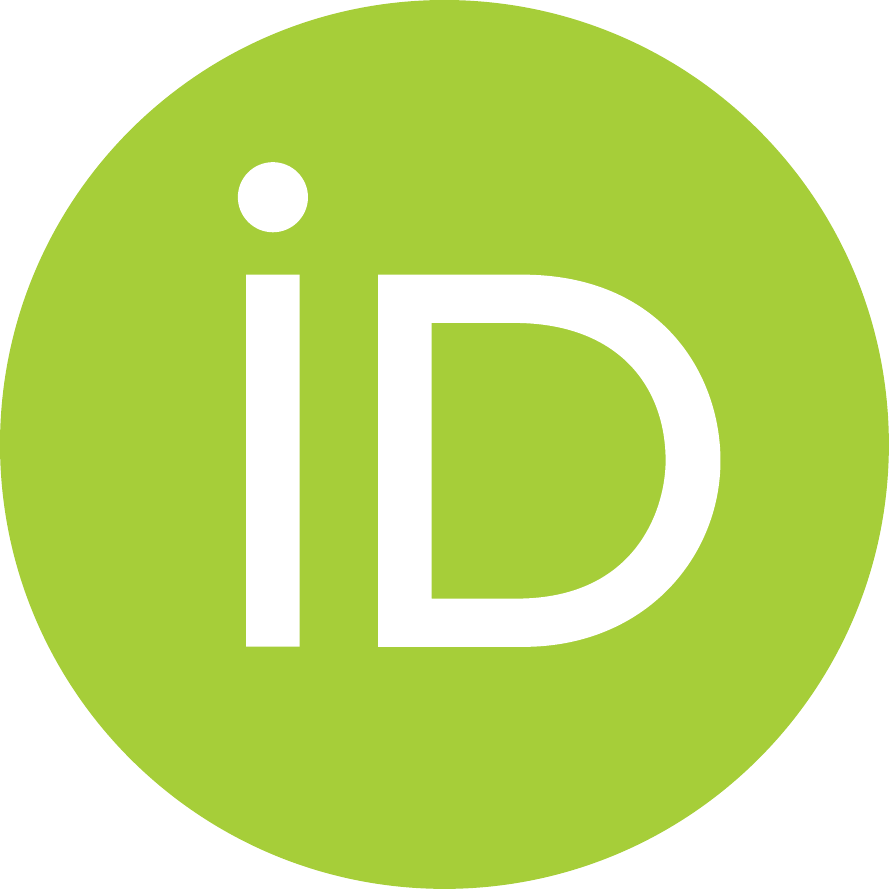}}}

\lstdefinestyle{myCustomMatlabStyle}{
    language=C++,
    basicstyle=\color{darkgray}\footnotesize\ttfamily,
    keywordstyle=\color{amaranth}\ttfamily,
    stringstyle=\color{amaranth}\ttfamily,
    commentstyle=\color{amber}\ttfamily,
    morecomment=[l][\color{magenta}]{\#},
    numberstyle=\small, 
    stepnumber=1
}

\begin{document}


\title{Stellar Mergers with HPX-Kokkos and SYCL: Methods of using \\ an Asynchronous Many-Task Runtime System with SYCL }



\author{Gregor Dai\ss\orcid{0000-0002-0989-5985} \\
University of Stuttgart \\
\AND
        Patrick Diehl\orcid{0000-0003-3922-8419}, Hartmut Kaiser\orcid{0000-0002-8712-2806}\\
        Louisiana State University
        \AND
        Dirk Pfl\"uger\orcid{0000-0002-4360-0212} \\
        University of Stuttgart
}






\maketitle

\begin{abstract}

Ranging from NVIDIA GPUs to AMD GPUs and Intel GPUs: Given the heterogeneity of
  available accelerator cards within current supercomputers, portability is a
  key aspect for modern HPC applications. In Octo-Tiger, an astrophysics
  application simulating binary star systems and stellar mergers, we rely on
  Kokkos and its various execution spaces for portable compute kernels. In
  turn, we use HPX, a distributed task-based runtime system, to coordinate
  kernel launches, CPU tasks, and communication. This combination allows us to
  have a fine interleaving between portable CPU/GPU computations and
  communication, enabling scalability on various supercomputers.

However, for HPX and Kokkos to work together optimally, we need to be able to
  treat Kokkos kernels as HPX tasks. Otherwise, instead of integrating
  asynchronous Kokkos kernel launches into HPX's task graph, we would have to
  actively wait for them with fence commands, which wastes CPU time better
  spent otherwise. Using an integration layer called HPX-Kokkos, treating
  Kokkos kernels as tasks already works for some Kokkos execution spaces (like
  the CUDA one), but not for others (like the SYCL one).

In this work, we started making Octo-Tiger and HPX itself compatible with SYCL.
  To do so, we introduce numerous software changes most notably an HPX-SYCL
  integration. This integration allows us to treat SYCL events as HPX tasks,
  which in turn allows us to better integrate Kokkos by extending the support
  of HPX-Kokkos to also fully support Kokkos' SYCL execution space.

We show two ways to implement this HPX-SYCL integration and test them using
  Octo-Tiger and its Kokkos kernels, on both an NVIDIA A100 and an AMD MI100.
  We find modest, yet noticeable, speedups ($1.11$x to $1.15$x for the relevant
  configurations) by enabling this integration, even when just running simple
  single-node scenarios with Octo-Tiger where communication and CPU utilization
  are not yet an issue. We further find that the integration using event polling
  within the HPX scheduler works far better than the alternative implementation
  using SYCL host tasks.
\end{abstract}


\cleardoublepage
\newpage
\twocolumn

\section{Introduction}

Modern GPU-supercomputers like ANL's Aurora, NERSC's Perlmutter and ORNL's
Frontier contain thousands of compute nodes each, but use different GPUs,
ranging from NVIDIA\textsuperscript{\textregistered} GPUs to
AMD\textsuperscript{\textregistered} GPUs and
Intel\textsuperscript{\textregistered} GPUs. Developing portable
High-Performance-Computing (HPC) applications for these kinds of systems
requires both compute kernels that can efficiently target all these different
GPUs (and ideally still work well on CPUs), as well as efficient communication
and work scheduling to avoid bottlenecks when scaling to more GPU nodes,
enabling distributed scalability.
In Octo-Tiger, an C\texttt{++} astrophysics code used to simulate binary star
systems and stellar mergers~\cite{marcello2021octo}, we address this by using a
combination of Kokkos~\cite{9485033} and the C\texttt{++} standard library for
parallelism and concurrency (HPX)~\cite{kaiser2020hpx}. With Kokkos, we can
write portable compute kernels running on both CPU and GPU thanks to the
various memory and execution spaces within. We further use HPX, a distributed
asynchronous many-task runtime system, to express dependencies between kernels
with HPX futures, call methods on remote compute nodes, overlap communication
and computation, and ultimately scale to thousands of compute nodes on machines
like CSCS's Piz Daint~\cite{10.1145/3295500.3356221} or ORNL's Summit
\cite{diehl2021octo}.

In previous work, we integrated Kokkos with HPX, allowing us to obtain HPX
futures for asynchronous Kokkos kernel launches and deep copies, thus
embedding them into the HPX task graph seamlessly~\cite{daiss2021beyond}. With this
integration, we do not need to explicitly synchronize the GPU when we need the
results on the host. Instead, we simply create a continuation for the
respective HPX futures, using the results automatically when they are available
(for instance, post-processing on the host or communication).

However, this HPX-Kokkos integration only works for some Kokkos execution
spaces (the CUDA\textsuperscript{\textregistered}, HIP and HPX execution spaces to be precise). For it to
function, there needs to be a deeper integration between HPX and the respective
underlying API,
allowing us to obtain an HPX future for asynchronous API calls (for example for \lstinline[language=c++]{cudaLaunchKernel}).

On unsupported execution spaces, the HPX-Kokkos integration has to wait (fence) for the
GPU results to arrive first and then return a ready dummy future (for
compatibility, even though at this point it is not asynchronous anymore). This
wastes CPU time, as the wait is actively blocking the CPU thread calling it
until the GPU kernel and associated memory copies are done, which in turn
delays other CPU tasks such as communication or working on more GPU kernel
launches. Notably, Kokkos already contains a
SYCL\textsuperscript{\texttrademark} execution space to support Intel GPUs,
however, it is not yet supported by HPX-Kokkos.

In this work, we address this by introducing an \textbf{HPX-SYCL} integration,
therefore extending the number of supported execution spaces of HPX-Kokkos by
the SYCL execution space and test it and its
performance with Octo-Tiger on both NVIDIA and
AMD GPUs (as we lack access to current Intel GPUs).

Of course, this integration also allows using SYCL and HPX together without
using Kokkos, as it makes the HPX task graph aware of asynchronous SYCL
operations. This allows a user to, for instance, define an entire graph of operations with
SYCL as usual, but in the end also to get one HPX future for the final result, which can be
used to asynchronously schedule communication, or arbitrary other CPU
tasks, with an HPX continuation.

We investigate two alternative methods of implementing this HPX-SYCL integration:
SYCL already contains a way to asynchronously schedule CPU work within its own
graph using host tasks depending on SYCL events. One
way to integrate HPX and SYCL is to directly use such a SYCL \lstinline[language=c++]{host_task} to set an associated HPX future to the ready
state once the events it depends on are completed. 
However, we also show an additional way of integrating the frameworks which
does not rely on these SYCL host tasks and instead uses
event polling implemented within the HPX scheduler, as we found that this
vastly outperforms the \lstinline[language=c++]{host_task} alternative.

Therefore, the contributions of this work are as follows:
\begin{enumerate}
\item We added an HPX-SYCL integration.
\item We made Octo-Tiger compatible with SYCL, using the Kokkos SYCL execution
  space and our HPX-SYCL integration. This required additional modifications in other repositories (and
    adaptations within Octo-Tiger itself).
\item We collected runtime data using Octo-Tiger on an AMD MI100 and an
  NVIDIA A100, using not only the SYCL
    execution space, but also the same Kokkos compute kernels running on the
    respective native execution spaces (CUDA and
    HIP), as well as their predecessor kernels which are implemented in pure
    CUDA and HIP. We collected some of this data
    with the HPX-SYCL integration turned on, and other data with it turned off.
\end{enumerate}

These contributions yield multiple benefits: Octo-Tiger users benefit from an
additional Kokkos execution space being fully available (asynchronous instead of
blocking), potentially allowing it to run on
Intel GPUs and Aurora in the future. HPX
application developers benefit from the HPX-SYCL integration because of the ability to
fully incorporate SYCL into their own applications (either by using the Kokkos SYCL
execution space and HPX-Kokkos or by using pure SYCL). Although combining HPX with
pure SYCL might seem like an odd combination at first, since both frameworks
include task-based capabilities, they can actually complement each other well,
with SYCL taking care of performance-portable compute kernels and the data
dependencies between them, and HPX taking care of synchronizing the results via
futures and distributing the application onto multiple compute nodes.

In turn, SYCL application developers seeking to add distributed capabilities to
their applications may benefit by having HPX as a viable alternative to pure
MPI (or frameworks like Celerity~\cite{thoman2022celerity}).

The remainder of the work is structured as follows: As we focus on Octo-Tiger
as a motivating example, we first introduce Octo-Tiger and its software
dependencies (notably HPX and Kokkos) as well as its execution model in the
next section. Subsequently, in Section~\ref{sec:integration}, we focus on SYCL
and the required software additions (notably the HPX-SYCL integration) to make
it work asynchronously with HPX, HPX-Kokkos and, thus, with
Octo-Tiger. We then benchmark and test this integration with Octo-Tiger in
Section~\ref{sec:results}. In Section~\ref{sec:related:work} we outline related
work. Finally, we finish with a conclusion and ideas for future work in
Section~\ref{sec:conclusion}.

\begin{figure}[t]
    \centering
    \includegraphics[width=1.00\linewidth,trim={0 1cm 0 1cm},clip]{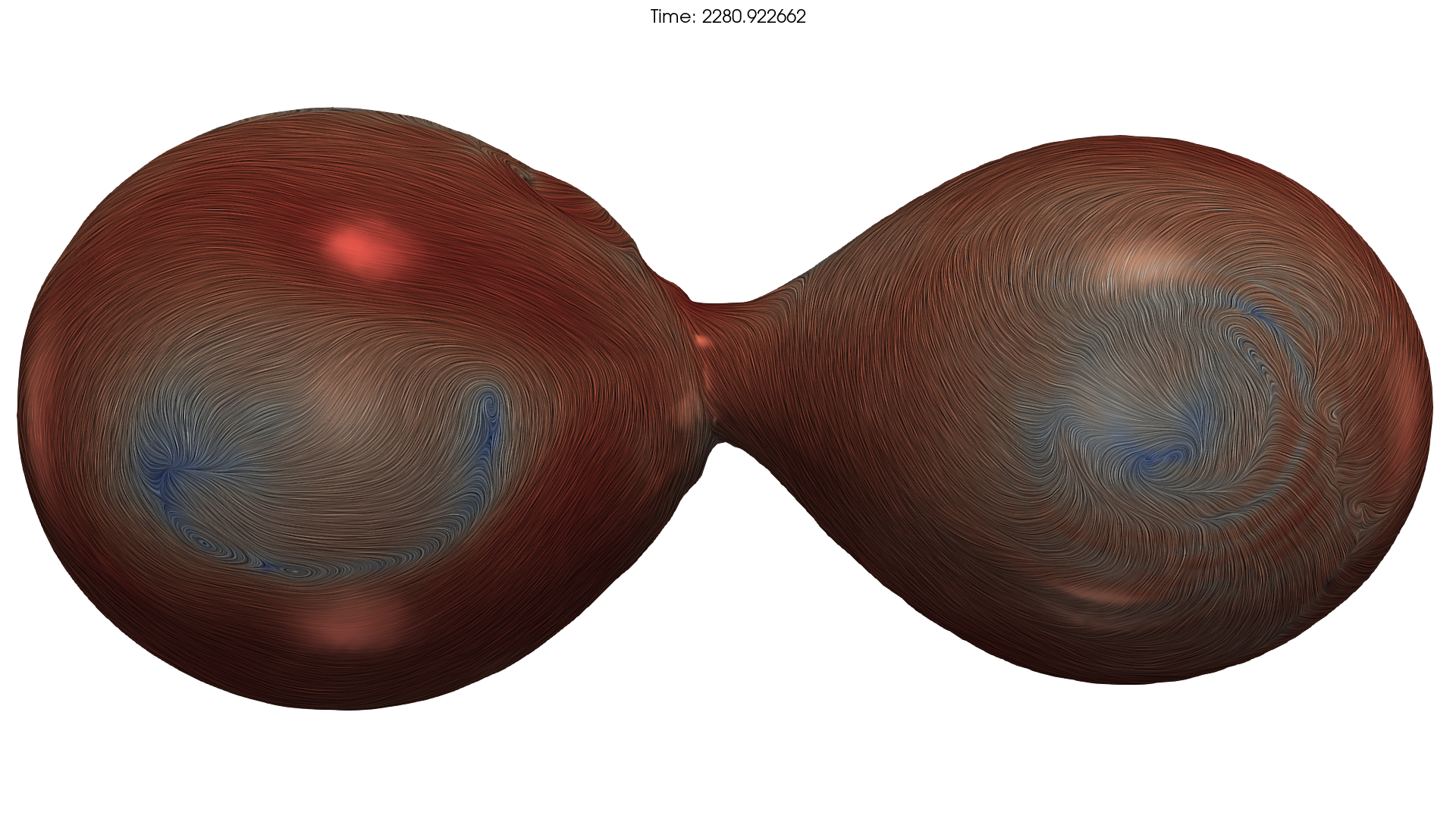}
    \caption{Snapshot of a double white dwarf merger with Octo-Tiger. Between the two stars, an aggregation belt is formed and mass from the smaller right star is transferred to the larger star on the left. The color shows the velocity magnitude of the stream, with red being high velocities.}
    \label{fig:octo_output}
\end{figure}
\section{Real-World Application as a Benchmark: Octo-Tiger}
\label{sec:octotiger}
The application driving our development efforts in this work is Octo-Tiger, as
we aim to run it on the Kokkos SYCL execution space. Octo-Tiger, itself, benefits
from the work shown here by gaining an additional execution space, allowing it
to target a wider range of platforms in the future.

Octo-Tiger is also an ideal benchmark to test integrations like the HPX-SYCL
one planned here, due to the tight interleaving of GPU and CPU work. Even in
small, single-node scenarios, we launch thousands of GPU compute kernels each
time step, tightly interleaved with CPU pre- and post-processing. 
Furthermore, it still contains CUDA and HIP
kernels for the main solvers, allowing us to compare performance using
multiple backends.
 
In the following, we introduce Octo-Tiger and briefly outline its solvers,
data structure and required software dependencies. Afterward, we describe its
execution model, based on all the required software, in more detail.

\subsection{Octo-Tiger: Scientific Application, Data-Structure and Solvers}
Octo-Tiger is a C\texttt{++} astrophysics code, used to study binary star
systems and their eventual outcomes, such as stellar
mergers~\cite{marcello2021octo}. These star systems contain two stars,
bound together by gravity. When close enough, they interact by slowly
exchanging mass. This can either be a stable transfer, occurring over millions
of years, or on unstable one which will disrupt one of the stars. Depending on
the system, this can lead to various outcomes, ranging from a Type Ia
supernovae, to the formation of another star. Figure~\ref{fig:octo_output}
shows a snapshot of a binary-star simulation done with Octo-Tiger at a point
where there's visible mass transfer occurring. In the past, Octo-Tiger was
already used to investigate R Coronae Borealis
stars~\cite{staff2018role} and the merger of bipolytropic
stars~\cite{kadam2018numerical}.

The stars are modeled as self-gravitating astrophysical fluids, using the
inviscid Navier-Stokes equation and Newtonian gravity. Hence, Octo-Tiger
contains two coupled solvers: A hydrodynamics solver using finite volumes and a
gravity solver~\cite{marcello2017very} using the fast-multipole-method (FMM).
These solvers operate on a 3D grid. Octo-Tiger uses adaptive mesh refinement,
maximizing the refinement for the area of interest, which usually is the
atmosphere between the stars. The data structure utilized for this is an
adaptive octree. For efficiency, instead of having just one cell per tree-node,
we use an entire sub-grid of cells per tree-node. The size of these sub-grids
is configurable at compile time, but the default we use is $8 \times 8 \times
8$ giving us $512$ per sub-grid (as larger sub-grids have adverse effects
regarding adaptivity, data-distribution, and FMM performance as the FMM uses the
tree-structure for approximations).

Highlighting its portability, Octo-Tiger was previously used on supercomputers
such as NERSC's Cori~\cite{heller2019harnessing}, Piz
Daint~\cite{10.1145/3295500.3356221} and ORNL's Summit~\cite{diehl2021octo}.
Recently, we began to target NERSC's Perlmutter and Riken's Supercomputer\
Fugaku.
It achieves this portability by using a mixture of HPX, Kokkos and other
frameworks as will be outlined in the next section.

\subsection{Octo-Tiger's Software-Stack}

\subsubsection{Tasks and Distributed Computing with HPX}
HPX is an Asynchronous Many-Task Runtime system (AMT)~\cite{kaiser2020hpx}. It is 
itself implemented in C\texttt{++}, and also implements all C\texttt{++}
20 APIs regarding parallelism and concurrency (including for example
functionality like \lstinline[language=c++]{hpx::mutex}). With HPX, we can
build an explicit task graph using HPX futures. Asynchronous operations return
futures, which can be chained together using continuations to form such a
graph. This way, subsequent work and communication is triggered automatically
whenever a task finishes. HPX also works in distributed scenarios, allowing us
to call functions on remote components asynchronously, getting futures in
return. 
Once their respective task dependencies are fulfilled, the HPX tasks within this task graph are executed by a pool of HPX worker
threads (each task by a single worker), with us usually using one worker thread per available CPU core. This means, while we may have
a lot of tasks available, we just have a few worker threads executing them over time.

Octo-Tiger is built entirely upon HPX, using the task-based programming framework
for efficient tree-traversals and for distributed computing. To this end, each
of the aforementioned sub-grids in the octree is a HPX component. This component contains
all required data for its sub-grid and communicates with other sub-grids using
remote function calls (via HPX) and HPX channels to fill its ghost layers and
coordinate computing. This makes each sub-grid as self-contained unit for
computing, easing distribution.

Consequently, all compute kernels only operate on one sub-grid (and its ghost
layers) at a time. However, we can usually invoke the compute kernels
concurrently for different sub-grids.

\subsubsection{Portable Compute Kernel with Kokkos and HPX-Kokkos} 
Octo-Tiger's compute kernels themselves were originally written for single-core
execution, with multiple kernels usually running at the same time for different
sub-grids. Over time, they were ported to support
Vc~\cite{https://doi.org/10.1002/spe.1149, Pfander18acceleratingFull}, CUDA and
HIP. To unify these separate kernels, we finally settled on Kokkos. Kokkos
implements a programming model for developing portable kernels~\cite{9485033}.
It contains multiple execution and memory spaces, allowing us to run kernels on
various devices (such as AMD and NVIDIA GPUs). Kokkos also already includes an
experimental SYCL execution and memory space, allowing its users to target Intel
GPUs. 

Moreover, Kokkos is tightly integrated with HPX: Both an HPX execution space
(using HPX worker threads to execute a Kokkos kernel) and an HPX-Kokkos
integration layer exist, the latter allows returning futures for Kokkos kernels
running on supported execution spaces (the
CUDA, HIP and HPX execution
space)~\cite{daiss2021beyond}.  With the HPX-SYCL integration introduced in
this work, the list of supported execution spaces now also includes the SYCL
execution space (as HPX-Kokkos directly utilizes the \lstinline{get_future}
functionality developed in this work).

\subsubsection{SIMD - Kokkos SIMD and \protect\lstinline[language=c++]{std::experimental::simd}} 
Kokkos kernels allow using explicit SIMD vectorization with C\texttt{++}
types~\cite{sahasrabudhe2019portable}. In our own kernels, we use this with the
Kokkos SIMD types\footnote{\url{https://github.com/kokkos/simd-math}},
enabling us to instantiate the types with, for example, AVX512 types when
compiling for CPUs and scalar double types when compiling for GPUs. The same
implementation also works using \lstinline[language=c++]{std::experimental::simd} on various CPUs. In previous
work, we investigated the performance difference between these two type libraries and
added SVE types~\cite{daiss2022simd} for Fujitsu A64FX\textsuperscript{\texttrademark} CPUs.

\subsubsection{Work Aggregation and Memory Optimizations - CPPuddle} 
Given the small, default size of the sub-grids with $512$ cells each, and the
fact that the compute kernels only work on one sub-grid at a time, the workload
per compute kernel is of essential importance for our GPU performance. 
Invoking just one such kernel is too little work to even come close to fully
utilize a GPU. As new sub-grids might be created over the duration of the
simulation and old ones may be deleted or migrated (all depending on the
adaptive mesh refinement as the simulation evolves), a static work aggregation
approach (defining sets of sub-grids to always be executed as one kernel) would
not work well for Octo-Tiger either.

Short of increasing the size of the sub-grids itself (negatively impacting
refinement), there are two ways of dealing with this: We can either rely on
executing enough GPU kernels concurrently (using multiple GPU executors) or
dynamically aggregate kernels as they are scheduled, depending on the load of
the GPU. For dynamic aggregation, we introduced work aggregation executors in
Octo-Tiger~\cite{daiss2022aggregation}. As these can be used in other HPX
applications, we extracted them into another software dependency, CPPuddle.
This library also contains memory pools for device-side buffers, which proved
to be handy when running the same task repeatedly for different sub-grids. If
the underlying GPU executor is currently busy, these aggregation executors will
start to bunch up compatible kernels (usually the same kernel but running on a
different sub-grid) as they are being scheduled. These will be launched as one
larger kernel when either an user-defined maximum number of aggregated kernels
is reached or the underlying GPU executor becomes idle and can work on it
immediately. For more details we refer to~\cite{daiss2022aggregation}.

\subsubsection{Other Dependencies}
Octo-Tiger uses some other dependencies, such as HDF5 and Silo, for the input
and output files. Dependencies like HWLOC and Boost are also indirectly
included in HPX.  Also, there are some older parts of Octo-Tiger that still use
Vc, however, we are in the process of removing those since we deprecated the Vc
support within Octo-Tiger in favor of using Kokkos kernels with explicit SIMD
types.

\subsection{Octo-Tiger's Execution Model}
\begin{figure}[t]
\centering
\includegraphics[width=.48\textwidth]{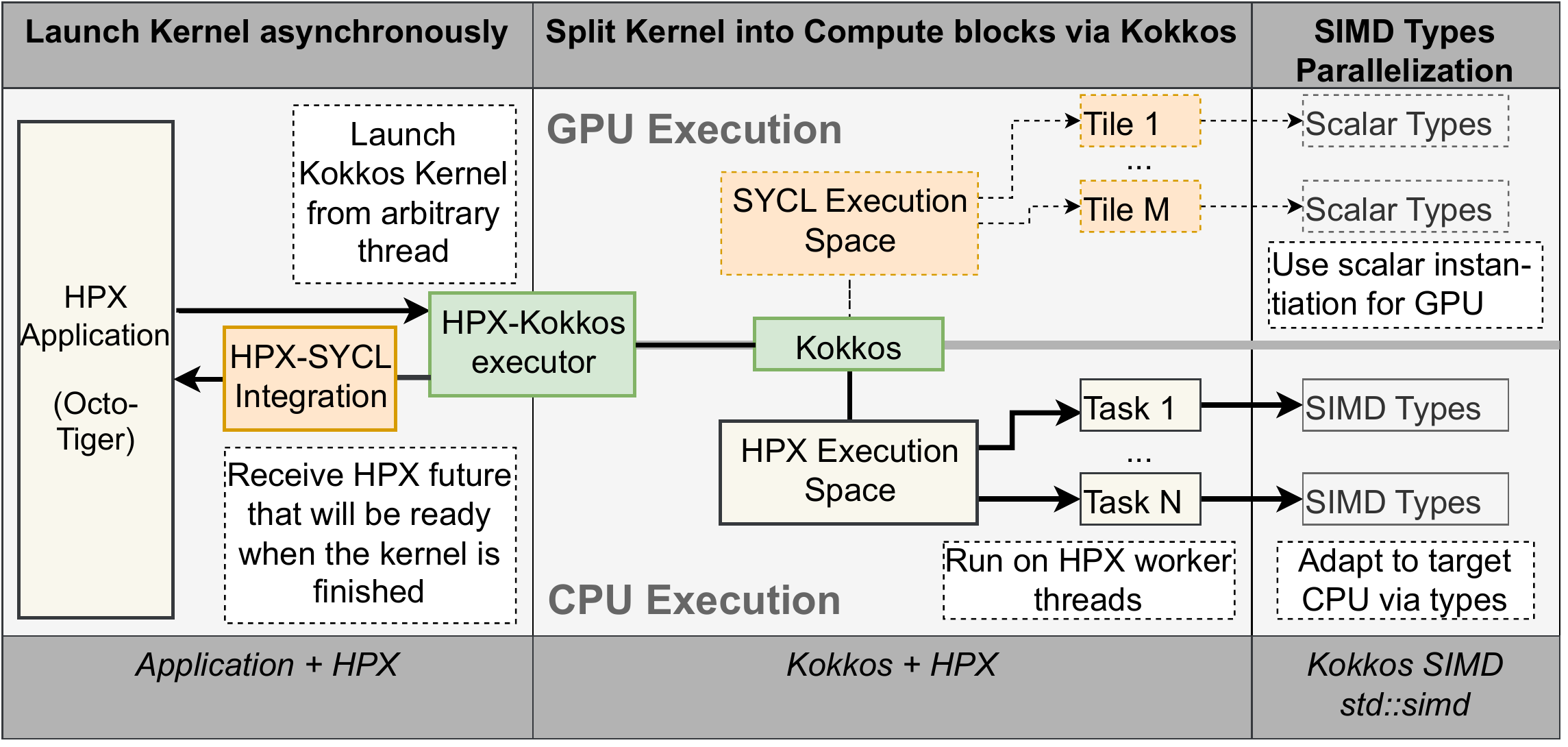}  
\caption{Octo-Tiger's execution model, adapted from~\cite{daiss2022simd} for the SYCL integration and associated execution space.}
\label{fig:model_sycl}
\end{figure}
During solver iterations, for instance with the gravity solver, Octo-Tiger traverses the
oct-tree that contains its grid-structure. To do so, we simply call the
methods executing the solver of neighboring (or child) tree-nodes. Thanks to
HPX, it does not matter whether they are actually located on the current
compute node or a remote node, as each tree-node is an HPX component.
Calling these methods via HPX returns futures, which we can use to chain
dependent tasks together. For example, we can first collect the data from
neighboring tree-nodes to fill the ghost layer of the current node, using the
associated future to chain an asynchronous Kokkos kernel launch.

In turn, the compute kernels work on the sub-grid of the current tree-node and
its ghost layer. The Kokkos kernel itself is launched asynchronously using
HPX-Kokkos, giving us yet another future. Each kernel is executed on
either a CPU execution space, usually the Kokkos HPX execution space which
splits the kernel into HPX tasks that will be executed by one or more of the
worker threads, or a GPU execution space, such as the Kokkos CUDA execution
space. Depending on the execution location, we instantiate the appropriate SIMD
types within the kernel, either with scalar types on GPUs or types using the
appropriate SIMD instructions on CPUs to make use of SIMD operations and masks.
We outline this execution model, and where to plug in the HPX-SYCL integration,
in Figure~\ref{fig:model_sycl}.
Of course, we could already use the Kokkos SYCL execution space without any further integration, albeit
only synchronously. However, to make the whole machinery described above work with
SYCL as it does with the supported execution spaces, we need HPX-Kokkos to be
able to return a future for the underlying SYCL kernel, hence we need the HPX-SYCL integration. Otherwise, the
asynchronous creation of the task graph using the futures would become
synchronous, impacting the overall performance. 

Lastly, also part of Octo-Tiger's execution model but not shown in this figure:
We usually use a pre-allocated pool of GPU executors, each being able to work
on GPU data-transfers/kernels independently. We can either use
only one of those (disabling concurrent GPU kernels altogether) or up to $128$
(theoretically more are possible but yield no further benefits). As working on
one sub-grid at a time might not be enough work to fully utilize some devices,
even given concurrent GPU kernels, we further employ the aforementioned dynamic
work aggregation by using the aggregation executors that can dynamically aggregate kernel
launches of the same kernel on different sub-grids. As shown in
\cite{daiss2022aggregation}, this is beneficial on
NVIDIA A100 GPUs and crucial on AMD MI100
GPUs.

\section{Integrating SYCL with HPX and Octo-Tiger}
\label{sec:integration}
In this section, we describe how we integrated SYCL with HPX, HPX-Kokkos, and
ultimately, Octo-Tiger. We start with a short SYCL introduction before
moving to the actual integrations in the following subsections. There, we first
cover the two variations of HPX-SYCL integrations that we tested and
afterward the changes to Octo-Tiger and its other dependencies.

\subsection{SYCL}
SYCL, a single source embedded domain-specific language (eDSL) aligned with the
C\texttt{++} 17 standard, provides high-level abstractions for various
acceleration cards and other devices. The SYCL specification is handled by the
Khronos group, and various implementations are available, for instance
ComputeCPP\textsuperscript{\texttrademark} which is developed by \textit{Codeplay
Software}.
\textit{hipSYCL} (currently in the process of being renamed into Open SYCL) is
developed at the University of Heidelberg~\cite{alpay2020sycl} and supports all
CPU architectures using OpenMP and Intel
GPUs using Level Zero; AMD GPUs using ROCm; and
NVIDIA GPUs using
CUDA. DPC\texttt{++} is part of
Intel OneAPI and supports Intel CPUs using OpenCL and
Intel GPUs using Level Zero or OpenCL; AMD
GPUs using AMD ROCm\textsuperscript{\texttrademark}; and NVIDIA GPUs using
CUDA. It is notable that hipSYCL and
DPC\texttt{++} are the only implementations supporting all three GPU
architectures. \textit{triSYCL} is developed by Xilinx and supports Intel and
AMD CPUs using OpenMP or Intel Thread Building Blocks (TBB) and ARM CPUs using
OpenMP. In addition, Xilinx FPGAs are supported using OpenCL.
For our purposes, we require at least a SYCL implementation that supports the
SYCL 2020 specification~\cite{sycl2020standard}. Specifically, we need support
for USM and \lstinline{in_order} queues. Support for SYCL host tasks is actually optional, as only one of our integration
implementations relies on it. As Kokkos uses various OneAPI-specific
extensions in its SYCL execution space, we focus on DPC++ in the rest of the paper.
However, we also tested the HPX-SYCL integration itself with hipSYCL.

\subsection{HPX-SYCL Integration}

%

\begin{figure*}[tb]
\begin{lstlisting}[caption={Using the basic HPX-SYCL integration.},label={lst:code1},deletekeywords={for}]
sycl::event event = queue.submit([&](sycl::handler& h) {
     /* insert SYCL dependencies */
     h.parallel_for(num_items, [=](auto i) {
                /* insert numeric code here */ });/
});
// Call HPX-SYCL integration
hpx::future<void> my_future =
   hpx::sycl::experimental::detail::get_future(event);
// Add task to be executed once the event is done
hpx::future<void> continuation_future =
   my_future.then([&continuation_triggered](auto&& fut) {
     /* insert CPU work/communication/post-processing */
   });
/* Suspend the current HPX task if kernel and continuation 
are not yet done. This does not block the worker thread,
it merely moves to work on another available task*/
continuation_future.get()
\end{lstlisting}
\end{figure*}

Here, we take a closer look at the HPX-SYCL integration. In its most basic
form, what we need is the functionality to create HPX futures from SYCL events.
If an event is not yet complete, the associated future should not be ready.
Once the event is completed, and thus all SYCL operations required for this
event are complete, the HPX future should eventually become ready, and thus
trigger all subsequent tasks depending on this future.  The creation of this
future needs to be low-overhead and completely non-blocking as otherwise we
lose the advantage gained by having asynchronous SYCL kernel launches.

There are two distinct ways to implement this: We can either use SYCL host tasks directly or implement an event polling scheme within the
HPX scheduler.

\subsubsection{Integration using SYCL Host Tasks}
\label{sec:integration:hosttasks}
SYCL itself allows creating host tasks that depend on SYCL events. Once the required
events enter the status \lstinline[language=c++]{info::event_command_status::complete}, the
\lstinline[language=c++]{host_task} is triggered and executed on the CPU by a thread managed by the SYCL
runtime. This SYCL \lstinline[language=c++]{host_task} may include arbitrary C\texttt{++} code, hence we can
use it to trigger an HPX future. Essentially, given a SYCL event, we simply
need to create a future that is not yet ready. By setting the internal future
data, we can set the future to the ready state, automatically triggering all
potential continuations defined by the user. Hence, we use the \lstinline[language=c++]{host_task} as a
sort of callback that will set this internal future data when triggered.

This is the simplest form of integration. However, it comes with the downside
of having to rely on the SYCL runtime to handle the execution of these
asynchronous host tasks efficiently. Therefore, this method may come with
great overhead (as all CPU cores should be busy with HPX worker threads
already). Nevertheless, we include this method here, as it seems to be the most
straightforward way to achieve the integration. However, in the results, we
show that (at least for Octo-Tiger) it is not a viable solution due to the
aforementioned overhead.

\subsubsection{Integration via Event Polling}

Instead of using SYCL host tasks, we can make use of HPX runtime itself here. As we
already have the HPX scheduler that is repeatedly called by worker threads in
between tasks (to get a new task), we can use it to poll SYCL events
periodically, and, depending on the execution status of the events, trigger
associated callbacks.  When we want a future for a given SYCL event, we simply
construct a future (not ready yet) and create a struct that contains the event
and a callback lambda that sets the data of the future when it is executed
(again, turning the future ready and triggering potential continuations). We add this \lstinline[language=c++]{event_callback} struct to the scheduler. To keep the
overhead low, we use an \lstinline[language=c++]{ConcurrentQueue} for this as
HPX already provides this data-structure. Using this, all threads can add their
own \lstinline[language=c++]{event_callbacks} for different events without
unnecessary locking.

\begin{figure}[t]
\centering
\includegraphics[width=.48\textwidth]{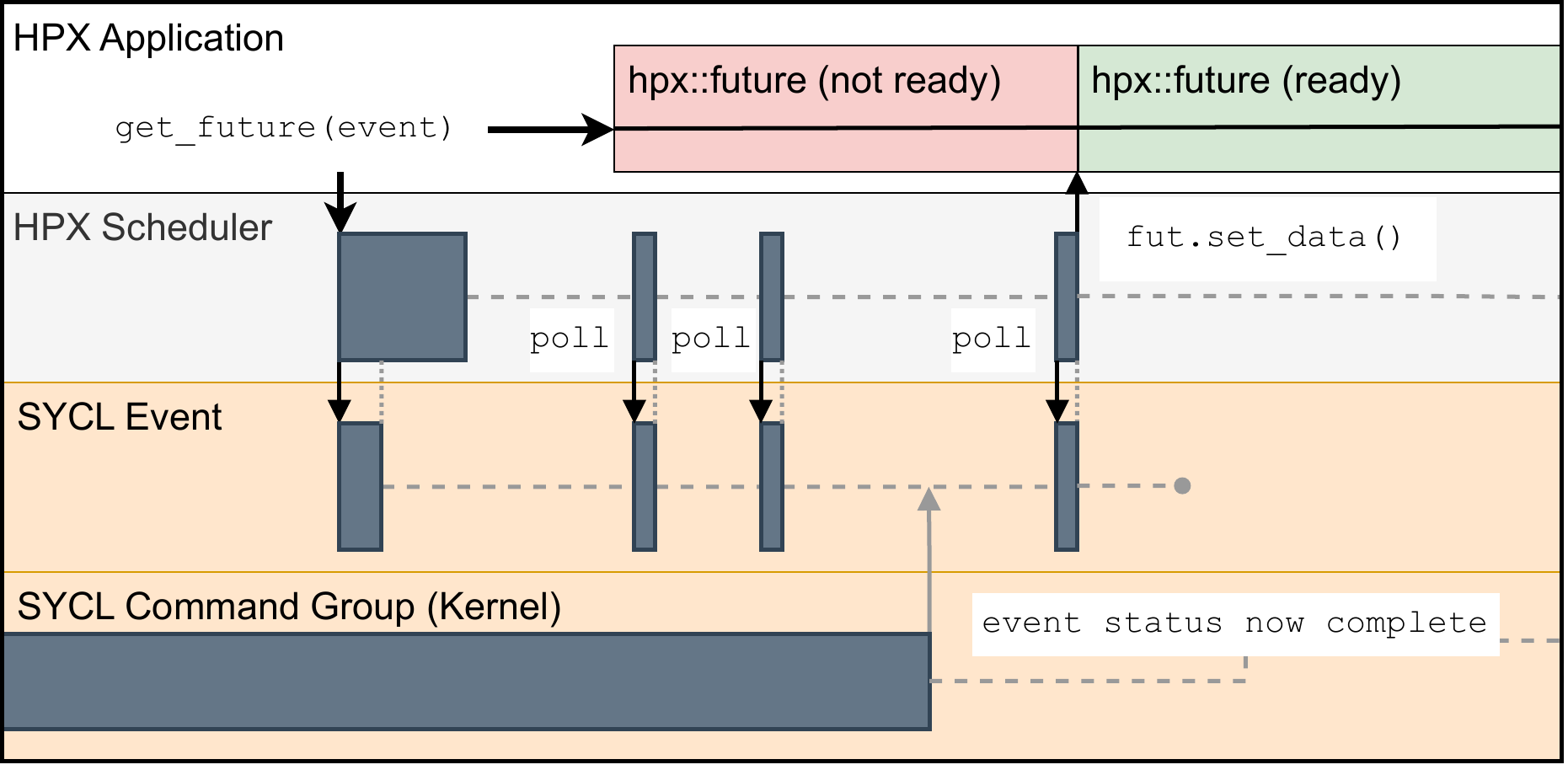}  
\caption{Outline of how the event polling works. The poll function is triggered by in between other tasks. The HPX future will become ready once the associated SYCL event is complete and the next poll detects it. Setting the future to ready will automatically trigger any potential HPX continuations.}
\label{fig:event_polling}
\end{figure}

The polling function that is called by the scheduler is a different matter:
This will be executed by only one thread at a time. That being said, other threads
trying to enter will not wait on the mutex, but instead return immediately, as
it is enough for one thread to do the polling at a time so other threads trying
to poll might as well work on other tasks (as there will always be another
thread visiting eventually as long as the HPX runtime is alive). The thread
executing the poll function tries to get all events from the concurrent queue
and checks each one for completion. If completed, the associated callback is
executed. If not, the \lstinline{event_callback} is moved to a vector to be polled again
at a later visit (inside the poll function, all events of the vector are polled
once as well). This way, an HPX future for a completed SYCL event will be
ready eventually, without ever having to call any blocking methods such as the event \lstinline[language=c++]{wait()} method. Instead, we merely have to poll
the events in-between other tasks which is handled automatically by our
integration inside the scheduler. This procedure is exemplified for one future
in Figure~\ref{fig:event_polling}. In Listing~\ref{lst:code1} is an example of
how this functionality can be used.

In case we need a future for an \lstinline{in_order} SYCL queue without having any SYCL
event available, we provide an overload for \lstinline{get_future} that inserts
a dummy SYCL \lstinline{single_task} and uses its SYCL event to get a future. It is
checked that the queue is \lstinline{in_order} as only in this case the dummy kernel will
be executed after all previous operations submitted to the queue, allowing us
to use this future to check if previously submitted commands are done even if
we do not have their events anymore (as can be the case when those commands are submitted
inside an external library).

We implemented this integration (see pull
request\footnote{\url{https://github.com/STEllAR-GROUP/hpx/pull/6085}}) and
initially
tested it with DPC\texttt{++} and with hipSYCL using some basic vector add and
stream examples. For it to work, the first hurdle was compiling HPX with the
respective SYCL implementation. For DPC\texttt{++}, we simply had to pass the
\lstinline[language=bash]{-fsycl} flag for the SYCL-related source files (such as tests and the
source files that implement the event polling) and \lstinline[language=bash]{-fno-sycl} for all other files. Our
initial attempt to do the same with hipSYCL's syclcc wrapper failed. Here, we
use hipSYCL's CMake integration instead, which worked out-of-the-box.

However, for the integration to work properly, the hipSYCL default
configuration needs to be considered. First, the hipSYCL runtime needs to be
kept alive. Without this, the entire SYCL runtime might be created and
destroyed inside the poll function if there are no other SYCL  objects alive at
this point, as the poll function creates a temporary \lstinline[language=c++]{event_callback}
(containing a SYCL event), which would cause re-creation of the hipSYCL
runtime.
Second, one must ensure the kernels are actually being launched: hipSYCL's
default scheduler supports automatic work distribution across multiple devices,
causing it to potentially delay launching a kernel since we will never wait on
either the event or the command queue.

Both issues can be addressed simply by properly configuring hipSYCL for this
usage, as the behavior can be oriented with the appropriate environment
variables. We can set the scheduler to the direct one, directly launching any
kernel/command, with \lstinline[language=bash]{HIPSYCL_RT_SCHEDULER=direct} and
make sure that the hipSYCL runtime remains alive with \lstinline[language=bash]{HIPSYCL_PERSISTENT_RUNTIME=1}. However, to be certain that the user
does not need to remember setting these, we added an extra command queue inside
the HPX scheduler, keeping the runtime alive as long as the HPX runtime itself
is alive. Furthermore, we can use this queue to trigger the flush method inside
hipSYCL to make the integration work with its default scheduler as well.

\subsubsection{HPX SYCL Executor}
\label{sec:hpx:sycl:executor}
On top of this basic \lstinline[language=c++]{get_future} integration, we added
a SYCL HPX executor. This executor wraps an \lstinline{in_order} SYCL command queue and
allows the dispatch of function calls by wrapping them in either \lstinline[language=c++]{hpx::async} (two-way execution, which means that it
returns an \lstinline[language=c++]{hpx::future}) or \lstinline[language=c++]{hpx::apply} (one-way execution without a future).
This executor only accepts SYCL queue member functions and passes them directly
to the underlying SYCL queue (eliminating the need to manually obtain the
future for these calls). This is done mostly for user convenience.
It also brings parity of features with the CUDA and HIP executors within HPX
that also support \lstinline[language=c++]{hpx::async} and \lstinline[language=c++]{hpx::apply}. In fact, we use an \lstinline{in_order} command
queue for the SYCL executor to keep the behavior the same as for these other
executors.

Like the basic integration described in the last section, this executor also
needed some adaptations when compiled with specific SYCL implementations. This
time, we needed to adapt it for DPC\texttt{++}, as picking the correct member
function overloads here requires taking into account the internal code location
parameter DPC\texttt{++} uses (requiring additional overloads taking this
parameter into account when compilation with DPC\texttt{++} is detected).

\subsubsection{Differences to other HPX GPU Executors} 
HPX also contains a CUDA executor (which
doubles as a HIP executor when compiled with hipcc).  This executor also
supports event polling (and, of course, \lstinline[language=c++]{hpx::async}
and \lstinline[language=c++]{hpx::apply}). It also supports using
CUDA callback functions directly. In fact, we
used this executor as a blueprint for implementing the SYCL executor and the
event polling SYCL integration.
However, there are some differences: In addition to getting everything compiled
with SYCL and changing the SYCL API calls, the SYCL  executor (and integration)
need to work with different SYCL implementations and thus need additional
logic to adapt to this (as with the aforementioned workarounds where we had
to adapt to hipSYCL or DPC\texttt{++}). Furthermore, we cannot implement an
event pool such as the one that the
CUDA executor uses, as SYCL itself lacks the
functionality to make use of our own events in calls (which would allow us to reuse
events, avoiding potentially expensive constructions and destructions). Here, we
have to simply use whatever events the SYCL queue is returning, and hope
the utilized SYCL implementation is using event pooling internally to reduce
the required overhead. On the one hand, this is beneficial when the utilized
SYCL runtime actually supports this, as it frees us from implementing it
ourselves: In fact, hipSYCL has event pooling for
CUDA and HIP events as of release 0.9.3.  On
the other hand, in the case where the runtime does not support event pools, we
have no way of adding our own within HPX.

\subsection{Additional Software Changes}
To get Octo-Tiger working with SYCL, we need more software additions.

\subsubsection{HPX-Kokkos}
First, we need to modify HPX-Kokkos itself (see pull
request\footnote{\url{https://github.com/STEllAR-GROUP/hpx-kokkos/pull/13}}). As
mentioned, HPX-Kokkos can return HPX futures for Kokkos kernel launches, but
only for supported execution spaces. Hence, we need to add the SYCL-execution
space to this list of supported spaces. This requires adding 
overloads, calling the correct \lstinline[language=c++]{get_future} method from
HPX instead of fencing and returning a dummy future which is the default behavior for
unsupported execution spaces. To do so, we need to hook into the basic
HPX-SYCL integration described in Section~\ref{sec:hpx:sycl:executor}, simply
by mapping the \lstinline[language=c++]{get_future} call here to the one
described there. We also need to correctly construct the HPX-Kokkos SYCL
executor instances. Not to be confused with the HPX-SYCL executor mentioned
above, such an executor is merely a wrapper around a Kokkos SYCL execution space.
To keep the behavior the same as for other Kokkos execution spaces, we use an
\lstinline{in_order} command queue to construct each SYCL execution space. Lastly, we add
an additional overload for \lstinline[language=c++]{deep_copy_async} which directly uses the
SYCL event from the memcpy to get the associated future.

\subsubsection{CPPuddle}
Furthermore, we need to modify CPPuddle (see pull
request\footnote{\url{https://github.com/SC-SGS/CPPuddle/pull/15}}). Octo-Tiger
uses CPPuddle to manage device memory pools to avoid unnecessary allocations by
reusing old allocations that are no longer in use. This is handy for repeated
GPU tasks that require similarly sized input buffers. Here, we merely need to
add the appropriate allocators to get the device/host memory pools using SYCL
USM mallocs.

\subsubsection{Octo-Tiger}
Finally, Octo-Tiger itself requires some modifications before being able to
work with SYCL (see pull request\footnote{\url{https://github.com/STEllAR-GROUP/octotiger/pull/432}}). Here, we
switch to the appropriate HPX-Kokkos executor (using the Kokkos SYCL execution
space underneath) and the CPPuddle allocators using SYCL USM memory. However, when
running Octo-Tiger's standard test suite, we still noticed the results being
off when running with SYCL. Using the SYCL math functions over the
std ones (\emph{i.e.}\ using \lstinline[language=c++]{sycl::sqrt} instead of \lstinline[language=c++]{std::sqrt}) fixed this issue.


\subsubsection{Kokkos}
\label{sec:integration:kokkos}
While not necessarily required to get Octo-Tiger working with SYCL, we also
experimented with some changes to Kokkos. First, since we are using \lstinline{in_order} queues,
we noticed that we can remove some barriers within the SYCL execution space in
Kokkos. Moreover, to use the Kokkos SYCL execution space on AMD GPUs, we
needed to make some modifications to the Kokkos CMake configuration to
correctly pass the arguments. These changes are not yet upstreamed as we want to refine and test them further first. However, both the patch removing
some of the fences is
available\footnote{\url{https://github.com/STEllAR-GROUP/OctoTigerBuildChain/blob/sycl_toolchain/kokkos_sycl_less_fencing.patch}}
as is the other
one\footnote{\url{https://github.com/STEllAR-GROUP/OctoTigerBuildChain/blob/sycl_toolchain_hip/kokkos_hip_arch.patch}}
enabling Kokkos builds with AMD GPUs.


\section{Results}
Here, we evaluate the performance impact of our software changes and
additions introduced in Section~\ref{sec:integration}. First, we introduce the
test setup, specifically we introduce the utilized Octo-Tiger scenario, hardware and software versions. 
Then we cover the parameters and patches used in our experiments and follow with the performance tests.
\label{sec:results}
\begin{figure*}[t]
\centering
\subfloat[\label{fig:integration-host-tasks-a100-1}A100: Increasing Number of GPU executors] {
\centering
  \includegraphics[width=.33\textwidth]{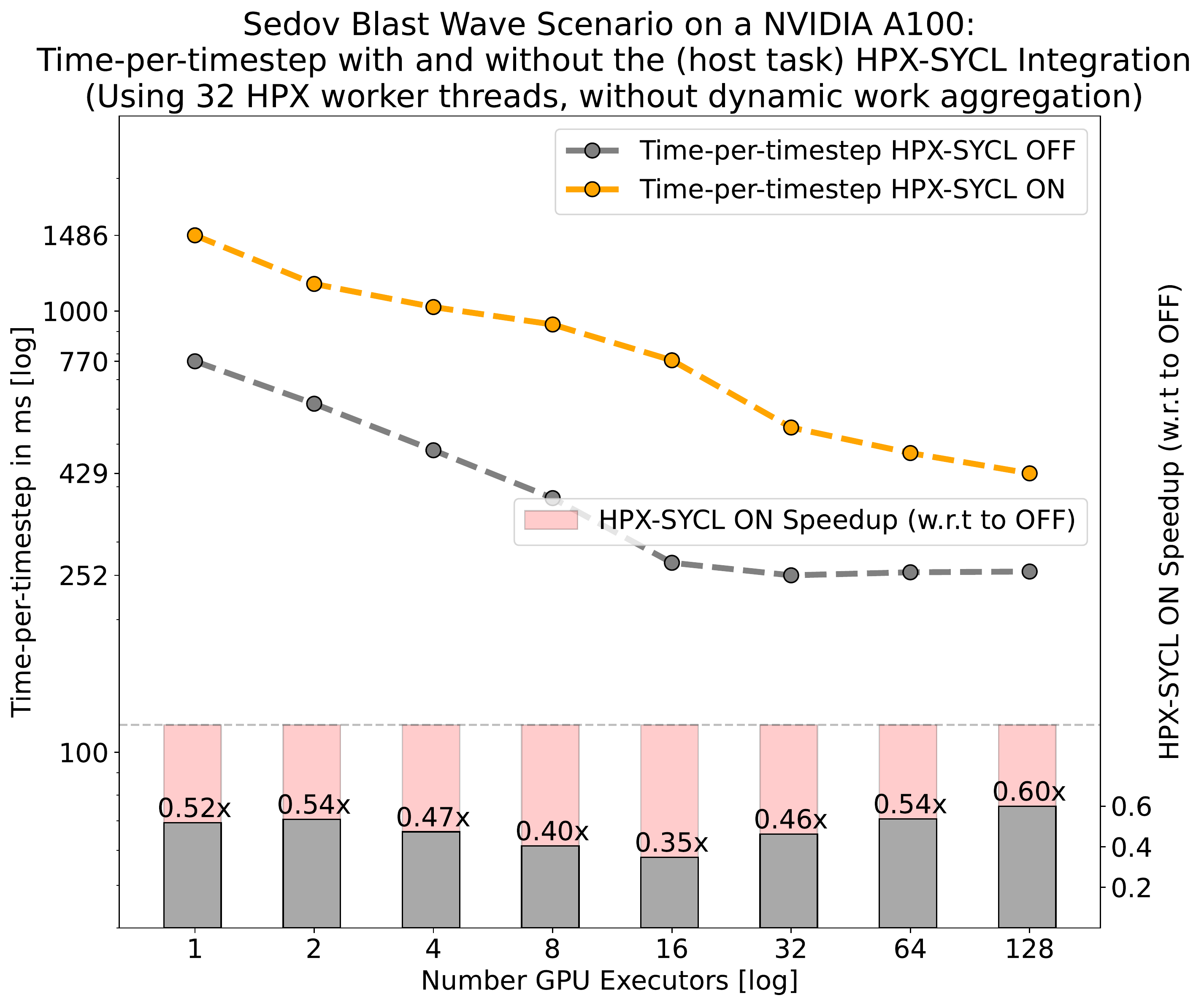}  
}
\hspace*{-0.1cm} 
\subfloat[\label{fig:integration-host-tasks-a100-2}A100: Increasing number of kernels aggregated] {
\centering
  \includegraphics[width=.33\textwidth]{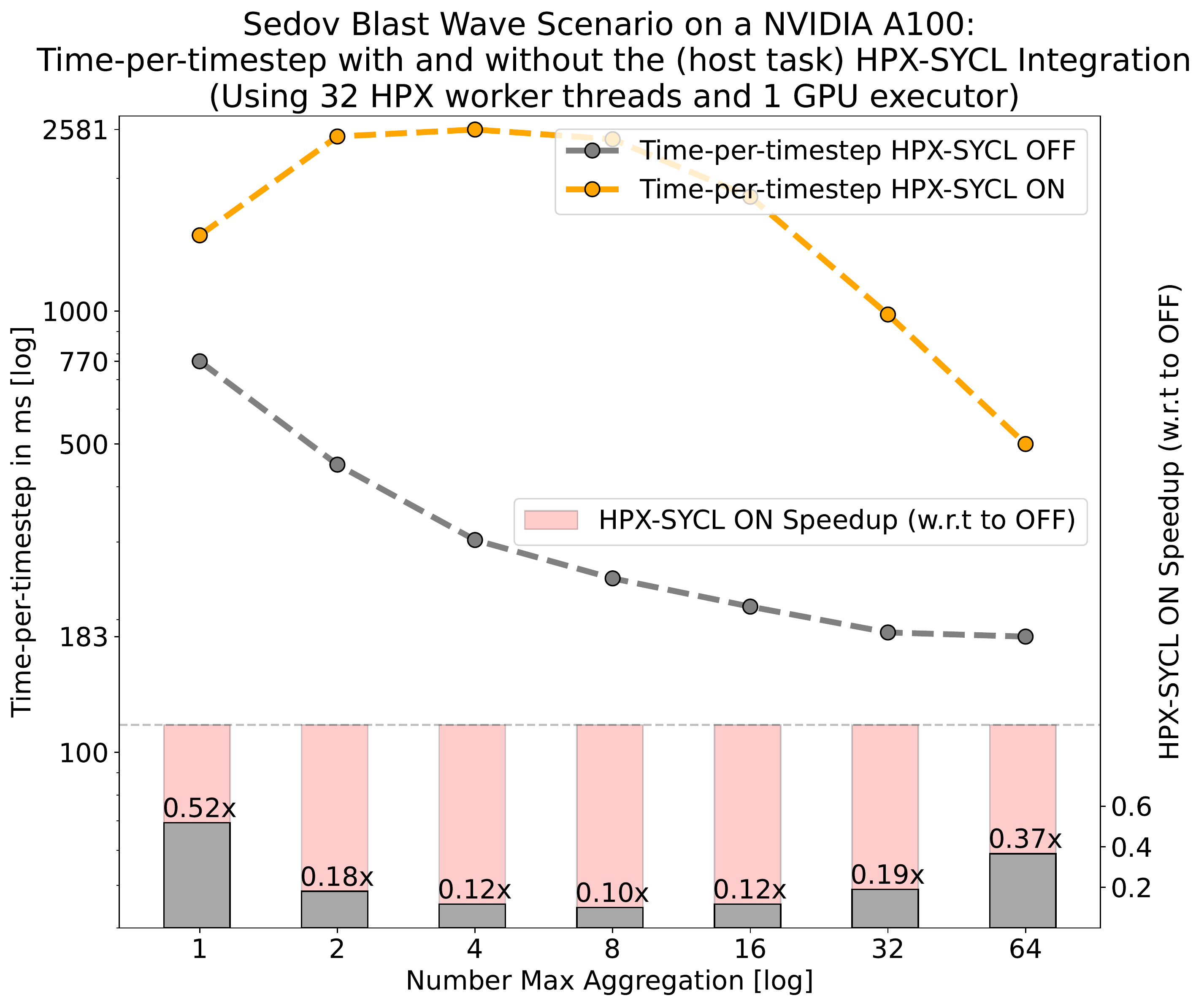}  
}
\hspace*{-0.1cm} 
\subfloat[\label{fig:integration-host-tasks-a100-3}A100: Best combinations] {
\centering
  \includegraphics[width=.33\textwidth]{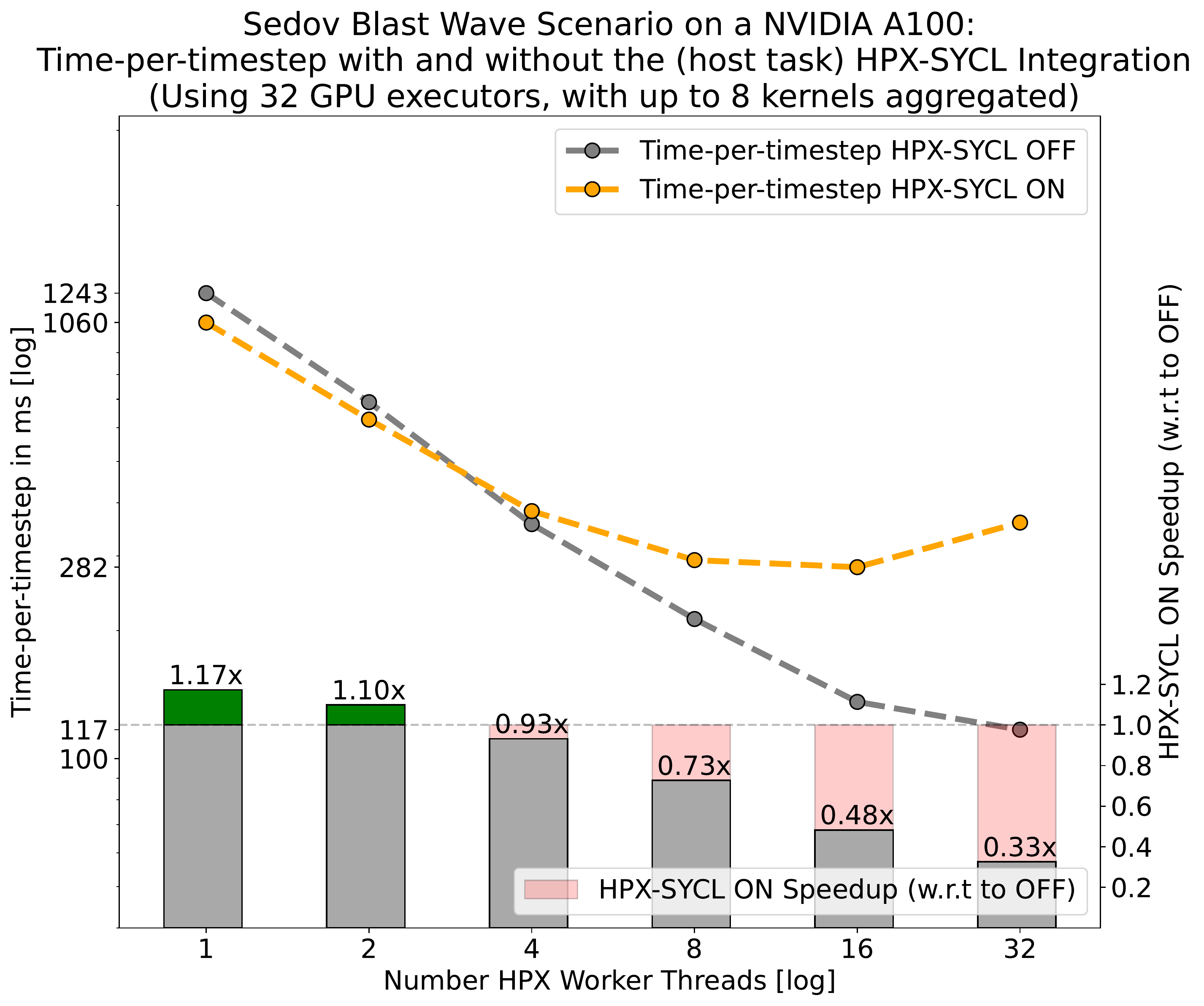}  
}
\hfill
\centering
\subfloat[\label{fig:integration-host-tasks-mi100-1}MI100: Increasing Number of GPU executors] {
\centering
  \includegraphics[width=.33\textwidth]{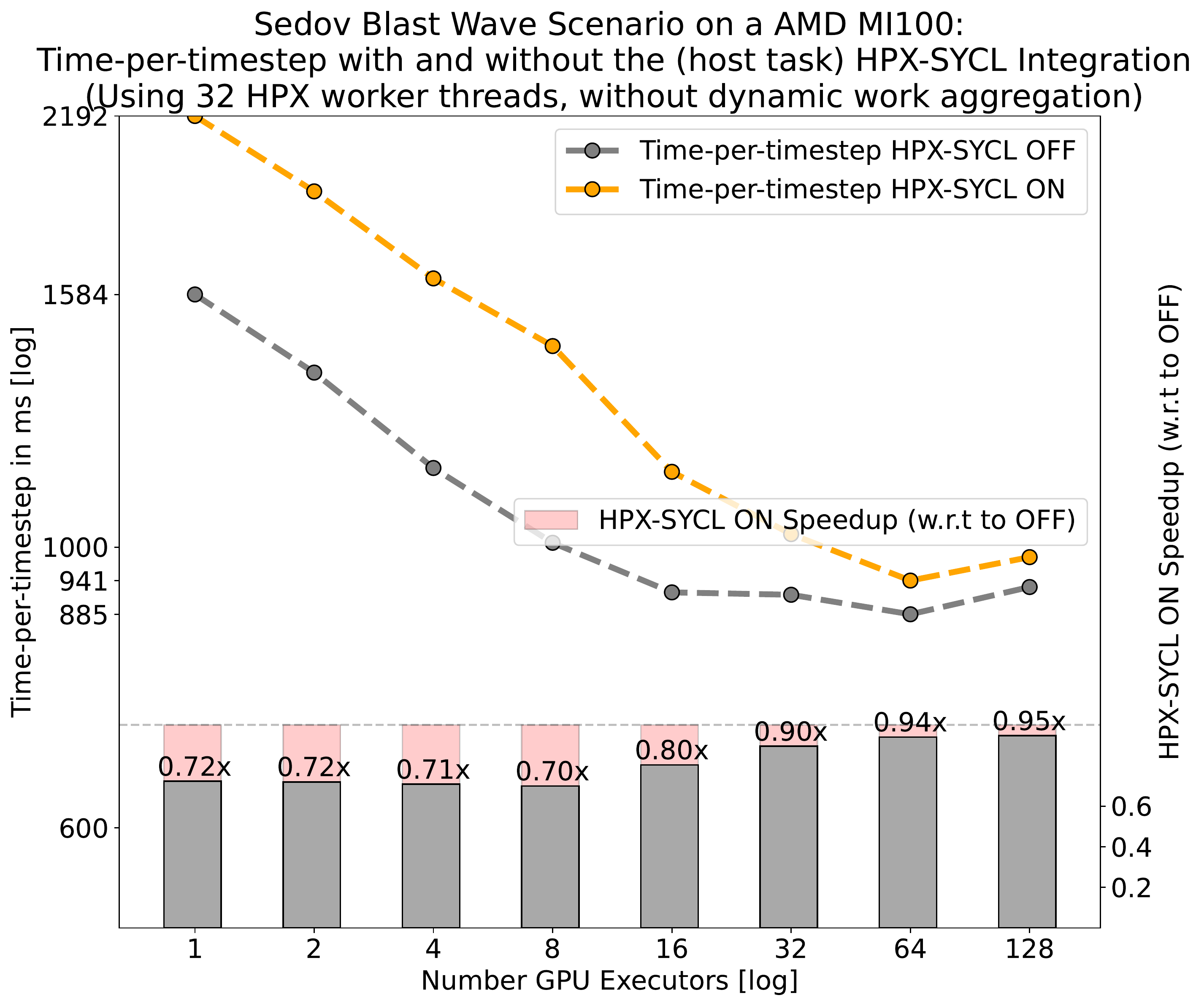}  
}
\hspace*{-0.1cm} 
\subfloat[\label{fig:integration-host-tasks-mi100-2}MI100: Increasing number of kernels aggregated] {
\centering
  \includegraphics[width=.33\textwidth]{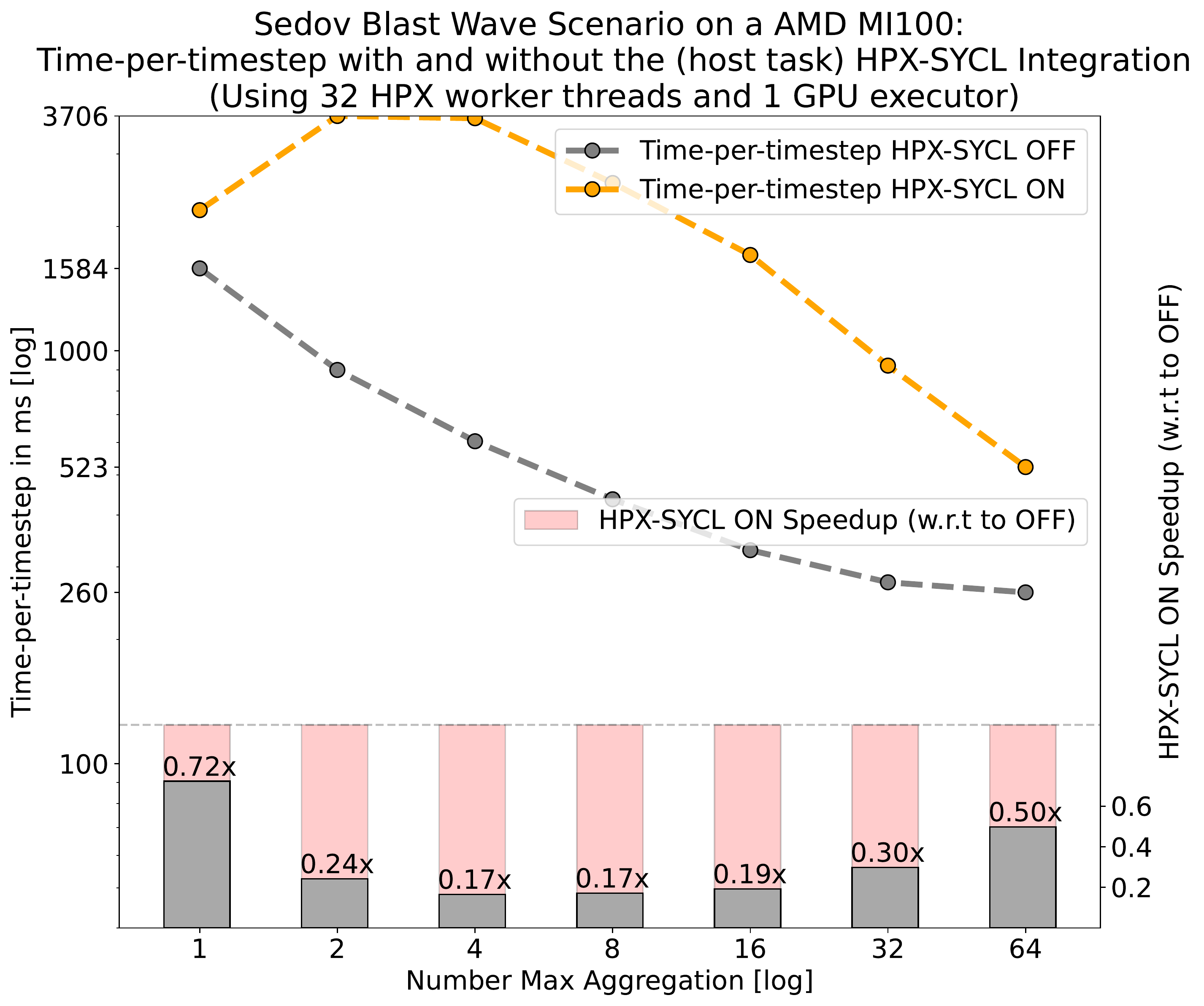}  
}
\hspace*{-0.1cm} 
\subfloat[\label{fig:integration-host-tasks-mi100-3}MI100: Best combinations] {
\centering
  \includegraphics[width=.33\textwidth]{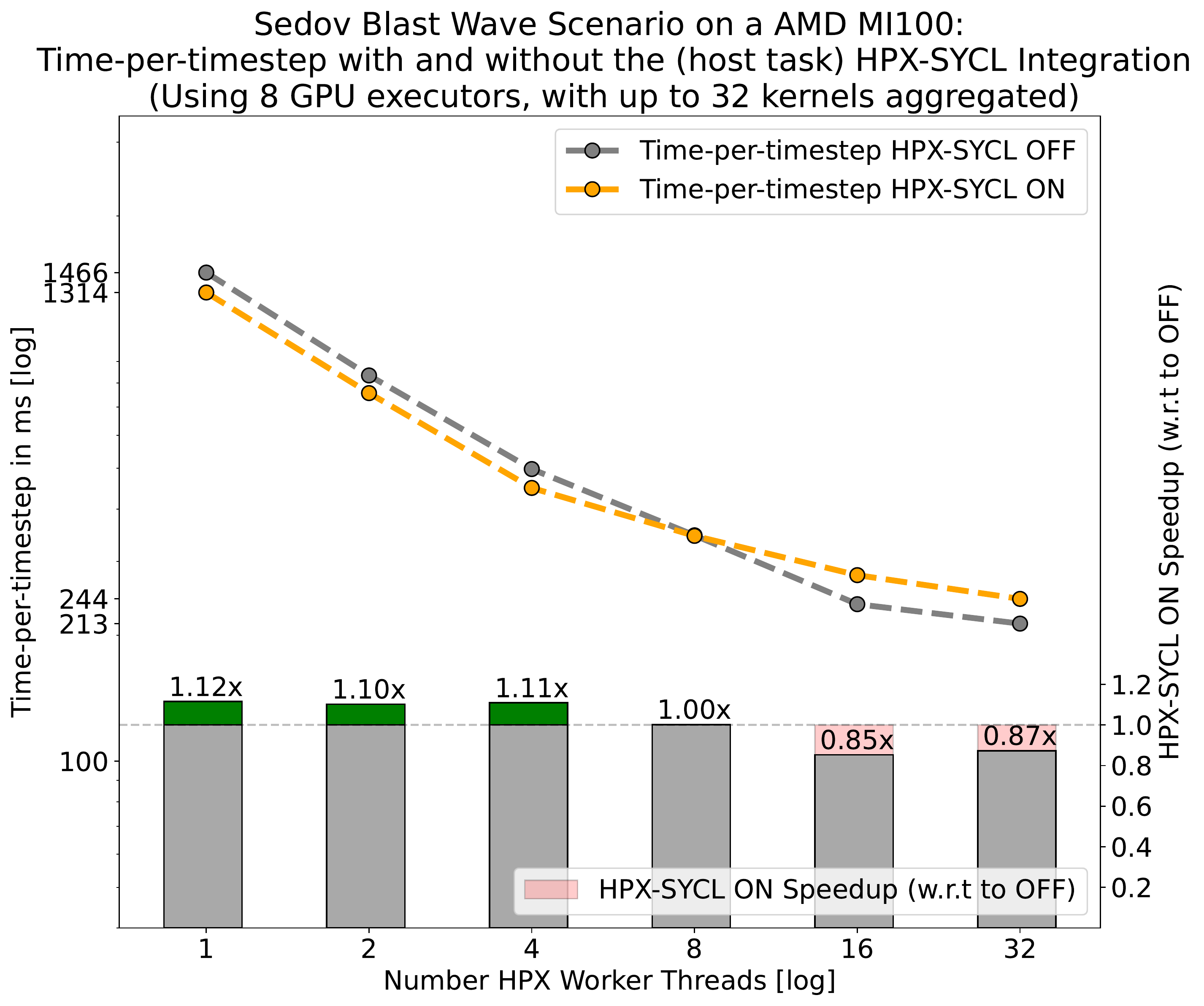}  
}
\caption{Runs with the SYCL-Integration (host task version) turned on (orange) and turned off (gray). In \ref{fig:integration-host-tasks-a100-1} and \ref{fig:integration-host-tasks-mi100-1} we do not use the dynamic work aggregation and instead only increase the number of GPU executors. In \ref{fig:integration-host-tasks-a100-2} and \ref{fig:integration-host-tasks-mi100-2} we only use one executor, but increase the maximum number of kernels aggregated into one kernel launch. In \ref{fig:integration-host-tasks-a100-3} and \ref{fig:integration-host-tasks-mi100-3} we see the best combinations. The Kokkos optimization patch is applied to all runs.}
\label{fig:integration-event-polling-host-taks}
\end{figure*}
\subsection{Test Setup}
\subsubsection{Scenario and Hardware:}
As a benchmark scenario, we choose the Sedov-Taylor blast wave. This scenario
is one of the benchmarks originally used to verify Octo-Tiger's output. It only
uses the hydrodynamic solver and has an analytical solution, making it ideal
for testing codes like Octo-Tiger.
For our purposes, being limited to the hydrodynamic solver is useful, since
this is the module that currently fully supports the dynamic work aggregation.
This allows us to finely tune the size of the compute kernels launched (by
bunching up sub-grids while the GPU executor is busy and launching them as one
kernel once said executor is done with previous work). Hence, we can more
easily take a look at the overheads of our integration for differently sized
kernels. The support within the gravity solver for this kind of work
aggregation is not yet complete, causing us to rely on the hydro-only scenario.

The computational effort required for this scenario is as follows: It
includes $512$ leaf sub-grids, resulting in $262144$ cells overall. Per
time-step we call 15 GPU kernels per leaf sub-grid, giving us 7680 GPU kernel calls
per time-step overall (with 15360 CPU-GPU data-transfers). This assumes we have
the dynamic work aggregation described above turned off and further highlights
why we need it within the hydro module in the first place. The measured runtime
per time-step is the average over $15$ time-steps, and we use double precision for
all simulations.
The best compute time reached for this exact scenario (on the same
hardware) in previous work was $86.6$ ms on an
NVIDIA\textsuperscript{\textregistered} A100~\cite{daiss2022aggregation}. This
runtime includes not just all GPU kernels, but also things like CPU-GPU
data-transfers, post-processing on the CPU, scheduling of the GPU kernels,
determination of the time-step size and notably, all logic required for the
work aggregation.
Hence, it is a scenario that tries to run a large amount of small kernels, with
the GPU busy running them and the CPU busy scheduling and potentially
aggregating them.

We run this scenario on two nodes: The first node contains a
NVIDIA A100 GPU, and an
Intel\textsuperscript{\textregistered} Xeon\textsuperscript{\textregistered}
Platinum $8358$ CPU. The second node contains an AMD MI100 GPU and an AMD
EPYC\textsuperscript{\texttrademark} 7H12 CPU. We use $32$ HPX worker threads on both CPUs to
keep the runs more comparable to each other (by thus limiting HPX to $32$ CPU cores on both machines).
The software versions (\textit{git commits}) we use can be found in Table~\ref{tab:software}.
In particular, we use DPC\texttt{++} (OneAPI) as a compiler and SYCL
implementation, since the Kokkos SYCL execution space makes use of OneAPI
extensions such as \lstinline[language=c++]{sycl_ext_oneapi_enqueue_barrier}
and \lstinline{sycl_ext_intel_usm_address_spaces}.

\begin{table*}[tb]
    \centering
    \begin{tabular}{llll} \toprule
     Boost &  1.75.0 & HWLOC & 2.7.1  \\
     HDF5 & 1.8.12  & Silo &  4.10.2  \\
     Kokkos & \textit{23a5e94}$^1$ & HPX-Kokkos & \textit{6f6b655}$^2$ \\
     JEMALLOC  & 5.2.1 & HPX & \textit{5641895}$^3$ \\
     CPPuddle & \textit{969902f}$^4$ & Octo-Tiger & \textit{e969470}$^5$ \\
     ROCm & 5.2.0 & CUDA & 11.7 \\
     DPC++/Intel OneAPI & \multicolumn{3}{l}{44c6437684d6} \\\bottomrule
    \end{tabular}
    \caption{Software versions used in the experiments. As we use multiple experimental pull requests, we added the exact git commits and associated PRs.}
    \begin{flushleft}
    \footnotesize{$1$ part of \url{https://github.com/kokkos/kokkos/pull/5628} } \\
    \footnotesize{$2$ part of  \url{https://github.com/STEllAR-GROUP/hpx-kokkos/pull/13}} \\
    \footnotesize{$3$ part of \url{ https://github.com/STEllAR-GROUP/hpx/pull/6085}} \\
    \footnotesize{$4$ part of \url{https://github.com/SC-SGS/CPPuddle/pull/15}} \\
    \footnotesize{$5$ part of \url{https://github.com/STEllAR-GROUP/octotiger/pull/432}}
    \end{flushleft}
   \label{tab:software}
\end{table*}

\subsubsection{Parameters and Configurations:}
We run Octo-Tiger in multiple software configurations by applying patches to
its dependencies.

The first
patch\footnote{\url{https://github.com/STEllAR-GROUP/OctoTigerBuildChain/blob/sycl_toolchain/remove_hpx_sycl_integration.patch}}
simply turns off the HPX-SYCL integration within HPX to allow us to judge its
benefits. To make everything compile correctly, the interface needs to stay the
same, hence return an HPX future for a given SYCL event. With the patch to turn
off the integration, HPX waits for the SYCL event and only then returns a ready
future via \lstinline[language=c++]{hpx::make_ready_future()}, effectively
turning it from an asynchronous operation to a synchronous one. By default, this
patch is not applied for the following tests, unless stated otherwise (HPX-SYCL
OFF).

The second patch contains the Kokkos optimizations mentioned previously. In
early tests of our integration we found that the Kokkos SYCL backend contains
multiple barriers (\lstinline{sycl_ext_oneapi_enqueue_barrier}), reducing the
benefits we gain with our integration. Fortunately, we also found that we can
get rid of some of those barriers when using \lstinline{in_order} queues, as this queue
property already enforces the same kind of ordering. By default, this patch is
applied for the following tests unless stated otherwise.

The third change we apply is switching between the \lstinline{host_task} based HPX-SYCL
integration and the event polling based integration. This is done at
compile time within HPX-Kokkos, as this is the point where we call the basic
\lstinline{get_future} functionality. Depending on the configuration here, we
either call the event polling version or the \lstinline{host_task} version.

Other parameters we consider are the number of HPX worker threads, the number of GPU
executors, and the maximum number of GPU kernels that may be aggregated
together by one executor. 
Firstly, the number of HPX worker threads defines how many overall CPU threads are
working on the available HPX tasks, thus this parameter effectively steers how many
CPU cores are used by HPX.
Secondly, the number of GPU executors steers the number of concurrent GPU
kernels/data-transfers that are possible. 
Lastly, the maximum number of aggregated GPU kernels requires a more detailed
explanation: When we encounter a kernel that is compatible for aggregation (as
marked by the programmer), we suspend the current task and wait for other
threads to hit the same kernel on different sub-grids. We then launch all of
them as one bigger, aggregated kernel if we either hit a maximum number of
tasks encountering the kernel (this is the maximum aggregation parameter), or
if the underlying GPU executor becomes idle which may trigger the aggregated
kernel to launch sooner (avoiding deadlocks caused by odd numbers
of sub-grids). 
 This aggregation increases the size of the actual kernels
 running on the GPU, thus avoiding starving the GPU with numerous but tiny compute kernels,
 and decreases the load on the GPU runtime as we need fewer overall API calls.

The dynamic work aggregation also influences how often the \lstinline[language=c++]{get_future} functionality is called. For each leaf
sub-grid we run five separate GPU kernels before we have to transfer the
results back to the host for post-processing and communicating them to the
neighbors.

With the work aggregation turned on (\emph{i.e.}\ a maximum larger
than one kernel), we have one additional \lstinline[language=c++]{get_future}
call at the beginning of the first sub-grid encountered when aggregating (as
this future is used internally by the aggregation executor to notice when the
GPU stream it uses becomes idle). However, this is only done once per
aggregated kernel.

The other \lstinline[language=c++]{get_future} call is to
communicate the results of the aggregated kernel. This means that we effectively decrease the number
calls to our HPX-SYCL integration when going beyond a maximum of two aggregated kernels.

\subsection{Performance Tests}
For our first two performance tests, we run Octo-Tiger with the
HPX-SYCL integration turned on (using host tasks for the first test and event polling for the second test), and then run it again with the HPX patch
 that disables the integration altogether for comparison. Afterward, we
look at the performance with and without our experimental Kokkos optimizations
mentioned in Section~\ref{sec:integration:kokkos}. Finally, we take a short
look at how the performance of these Kokkos SYCL runs relate to the same runs
using the Kokkos CUDA/HIP execution spaces and the native
CUDA/HIP kernels still within Octo-Tiger.
\subsubsection{Test 1 - Performance Impact of the Host Task Based HPX-SYCL Integration}

For our first test, we take a look at the \lstinline{host_task} version of our integration
as described in Section~\ref{sec:integration:hosttasks}.  We run the Sedov
Blast Wave scenario with and without the integration turned on. The results can
be found in Figure~\ref{fig:integration-event-polling-host-taks}. As we will do
in the following tests, we usually look at three different configurations: For
the first graph (\ref{fig:integration-host-tasks-a100-1}), the number of HPX worker threads (and thus CPU cores used
by HPX) were fixed to $32$ and the dynamic work aggregation was disabled. We
then increase the number of GPU executors until we reach $128$. This way we can
see the effects of multiple CPU cores trying to use the integration on a
varying number of SYCL command queues (as each GPU executor contains one
underlying \lstinline{in_order} SYCL command queue).

In the second graph (\ref{fig:integration-host-tasks-a100-2}), we use only one GPU executor, but enable the dynamic
work aggregation. This allows the aggregation of up to $64$ kernels into one
larger aggregated GPU kernel. Although, usually the number of kernels being
aggregated into a single kernel ends up being smaller than this maximum number,
as the aggregated kernel is being launched immediately when the executor
becomes idle (even if it has not reached maximum aggregation capacity yet).

Lastly, in the third graph (\ref{fig:integration-host-tasks-a100-3}), we take a look at the best combination of the
previous two parameters that we found ($32$ GPU executors with up to eight
kernels aggregated on the A100). We then run this combination not only with
$32$ HPX worker threads but also try fewer threads. This is usually the most interesting
test, as it not only tests the configuration we are most likely to use in
production runs, but also begins to artificially weaken the CPU performance (by using less
workers), which in turn should make the HPX-SYCL integration more valuable, as
waiting on SYCL results (with the integration turned off) blocks the CPU
threads.
The other three graphs (\ref{fig:integration-host-tasks-mi100-1}, \ref{fig:integration-host-tasks-mi100-2}, \ref{fig:integration-host-tasks-mi100-3}) work accordingly for the MI100 node, showing
the same experiments done for this machine. 
This part of the experiment setup
 will stay the same for the next two tests as well. While we change the software
 configuration for those tests, the parameters we vary stay the same.

Overall, looking at the speedup bars in each of the graphs in test 1, it becomes apparent
that the integration actually significantly decreases performance, most likely
due to the host tasks being handled by different threads internally by the SYCL
runtime.
This can be seen in two ways: We get better (integration) speedups when using less HPX worker threads 
in graphs \ref{fig:integration-host-tasks-a100-3} and \ref{fig:integration-host-tasks-mi100-3} as there is less contention between the HPX worker threads and the threads of the SYCL runtime.
We further see a more severe slowdown in scenarios which rely heavily on the dynamic work aggregation (\ref{fig:integration-host-tasks-a100-2}, \ref{fig:integration-host-tasks-mi100-2}), 
as the CPU has to manage the work aggregation scheduling here as well.
Interestingly, the run with the best combination of GPU executors and work
aggregation on the MI100 (\ref{fig:integration-host-tasks-mi100-3}), performs better than the one on the A100 (\ref{fig:integration-host-tasks-a100-3}) when
the integration is turned on. We plan to investigate this further in future
work. However, even here this integration is not too beneficial: We only see a
benefit when using few HPX worker threads (and thus few cores).
In most
configurations, this \lstinline{host_task} integration is detrimental to the performance.

\begin{figure*}[t]
\centering
\subfloat[\label{fig:integration-event-polling-a100-1}A100: Increasing Number of GPU executors] {
\centering
  \includegraphics[width=.33\textwidth]{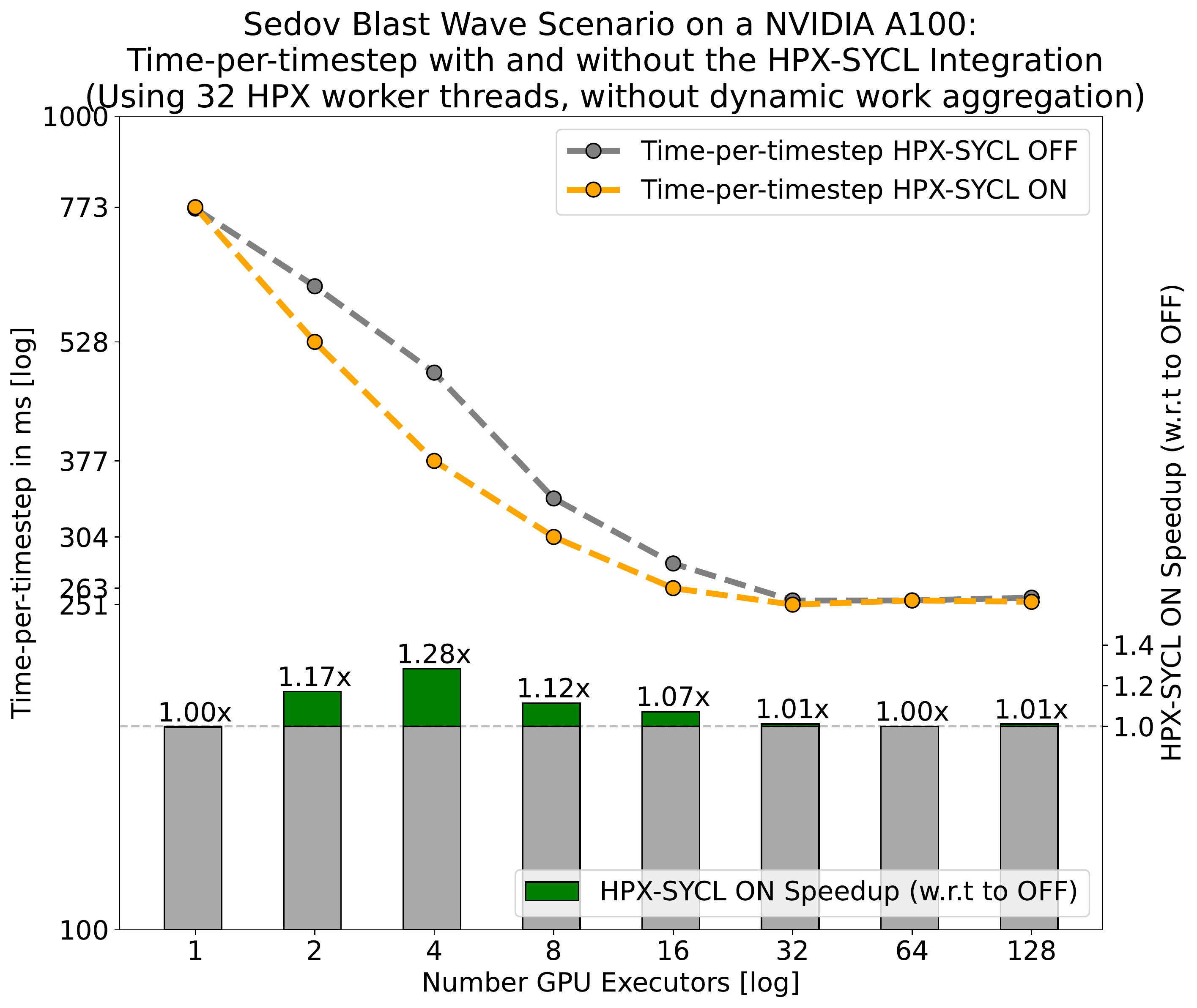}  
}
\hspace*{-0.1cm} 
\subfloat[\label{fig:integration-event-polling-a100-2}A100: Increasing number of kernels aggregated] {
\centering
  \includegraphics[width=.33\textwidth]{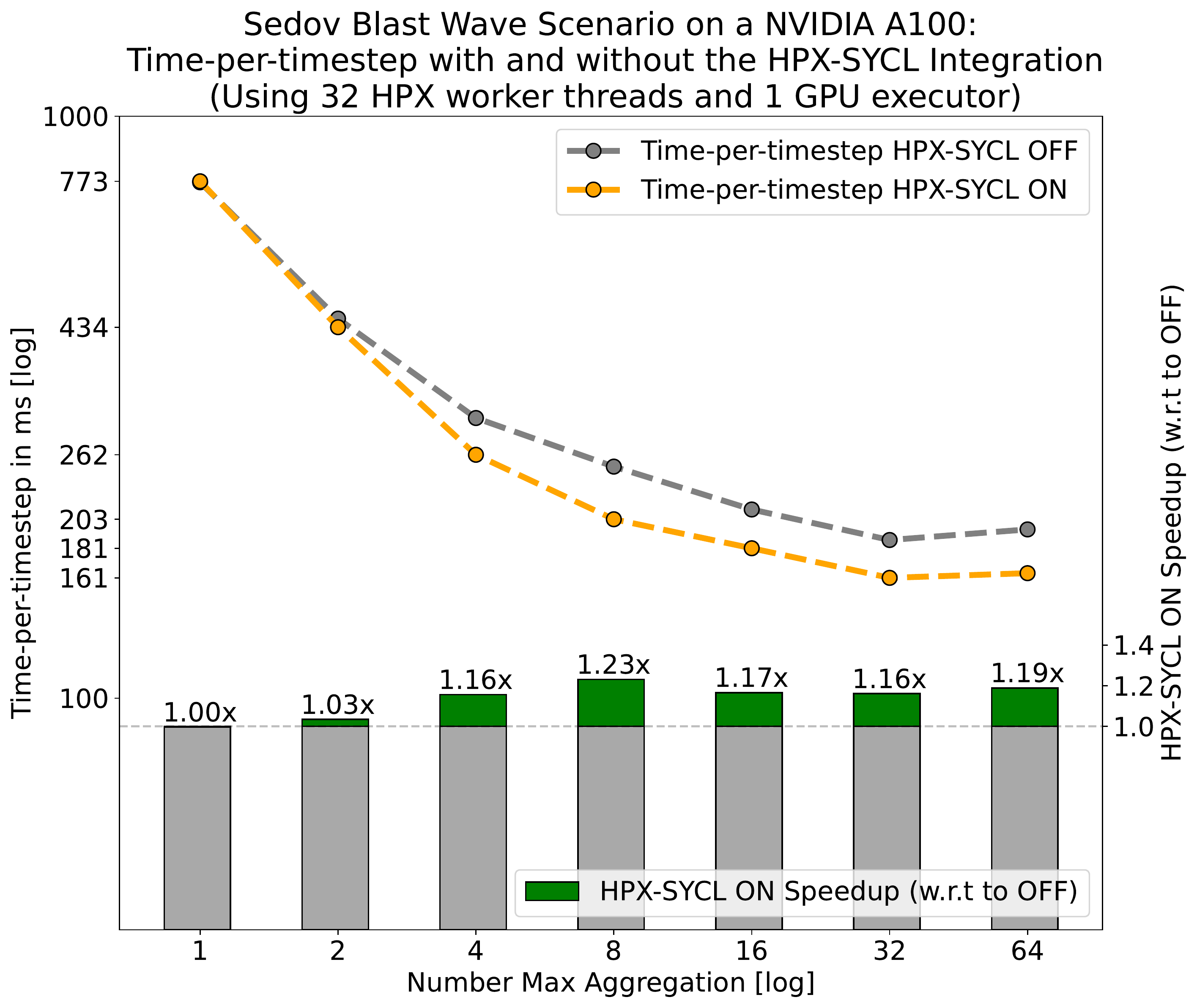}  
}
\hspace*{-0.1cm} 
\subfloat[\label{fig:integration-event-polling-a100-3}A100: Best combinations] {
\centering
  \includegraphics[width=.33\textwidth]{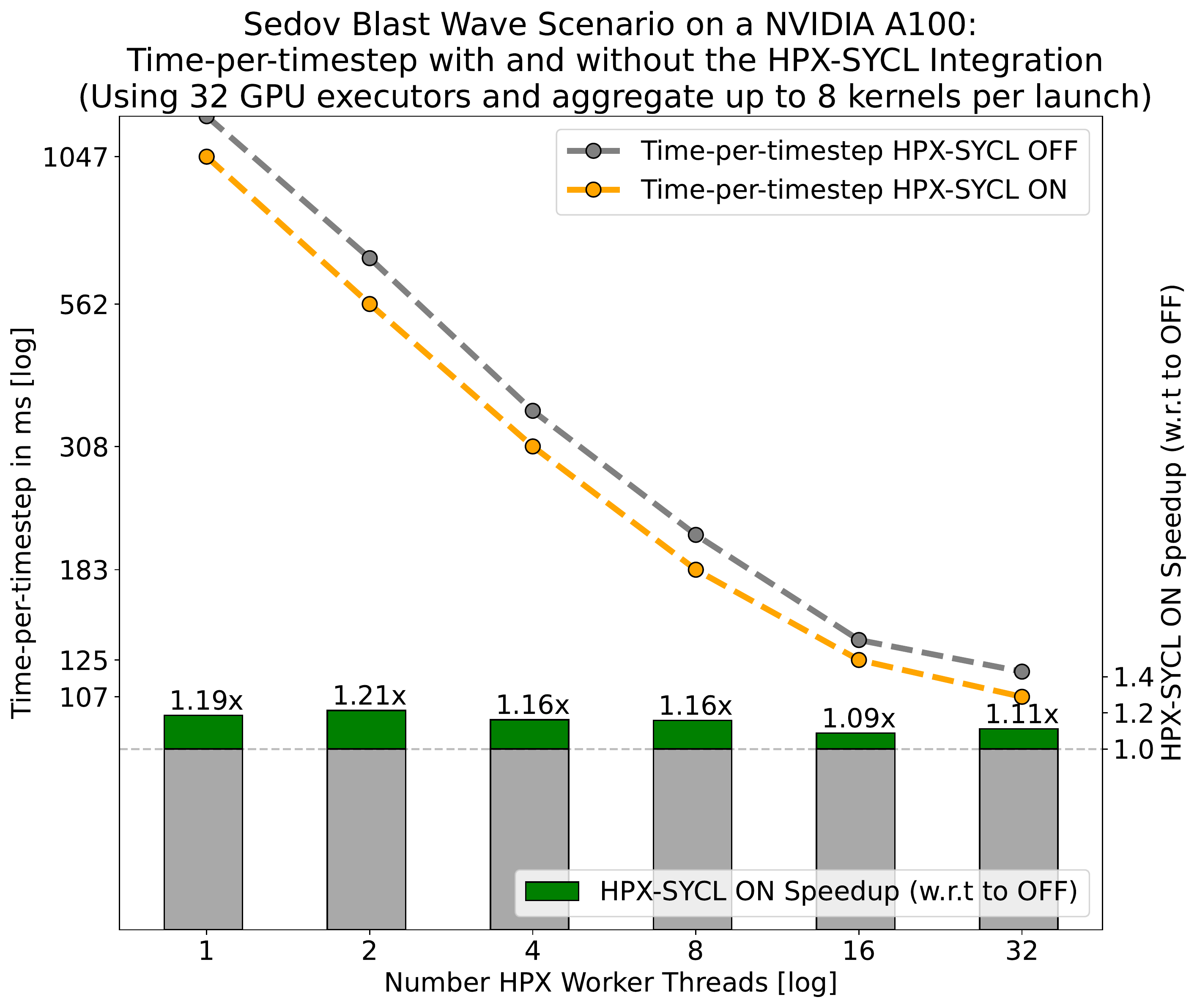}  
}
\hfill
\subfloat[\label{fig:integration-event-polling-mi100-1}MI100: Increasing Number of GPU executors] {
\centering
  \includegraphics[width=.33\textwidth]{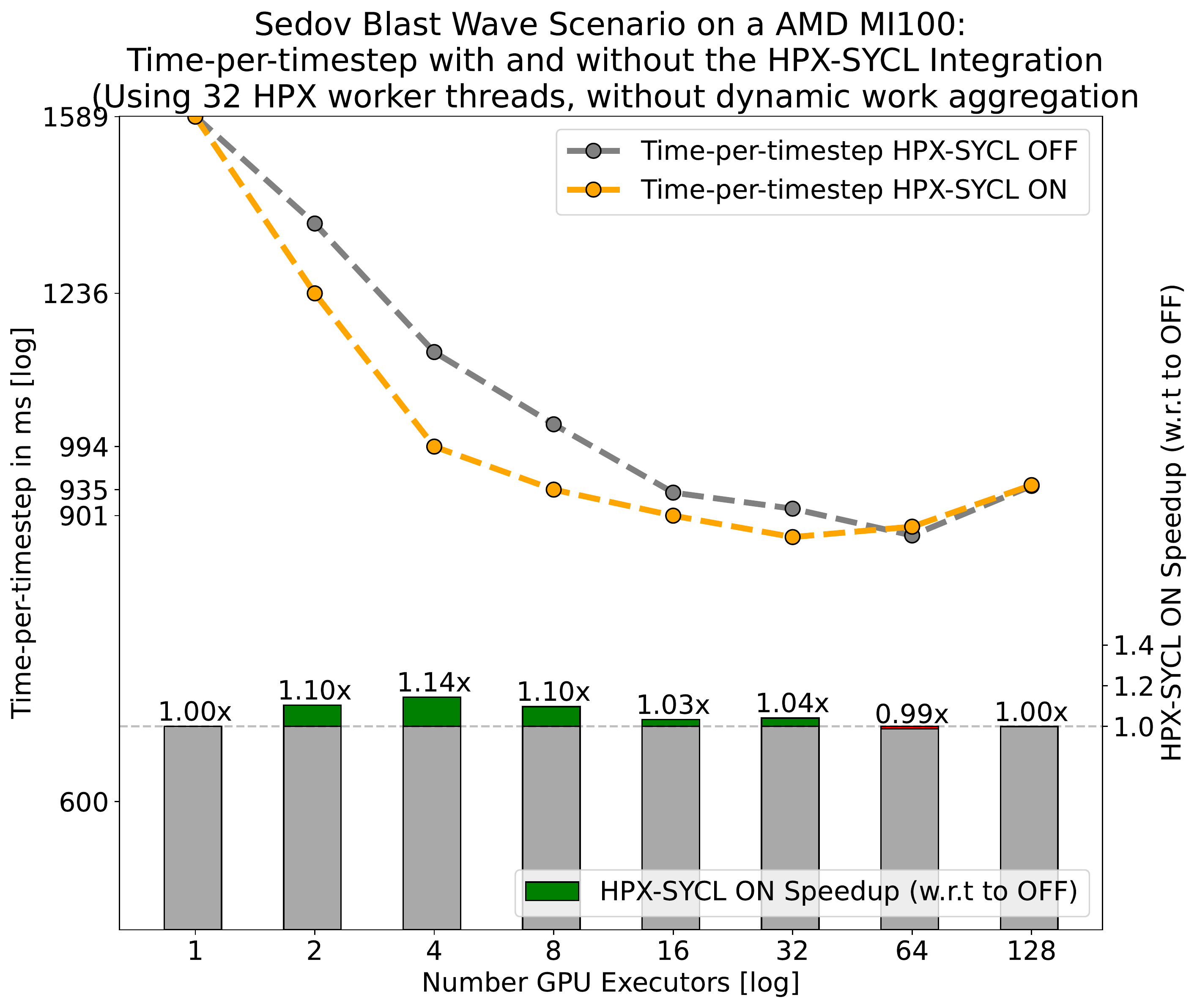}  
}
\hspace*{-0.1cm} 
\subfloat[\label{fig:integration-event-polling-mi100-2}MI100: Increasing number of kernels aggregated] {
\centering
  \includegraphics[width=.33\textwidth]{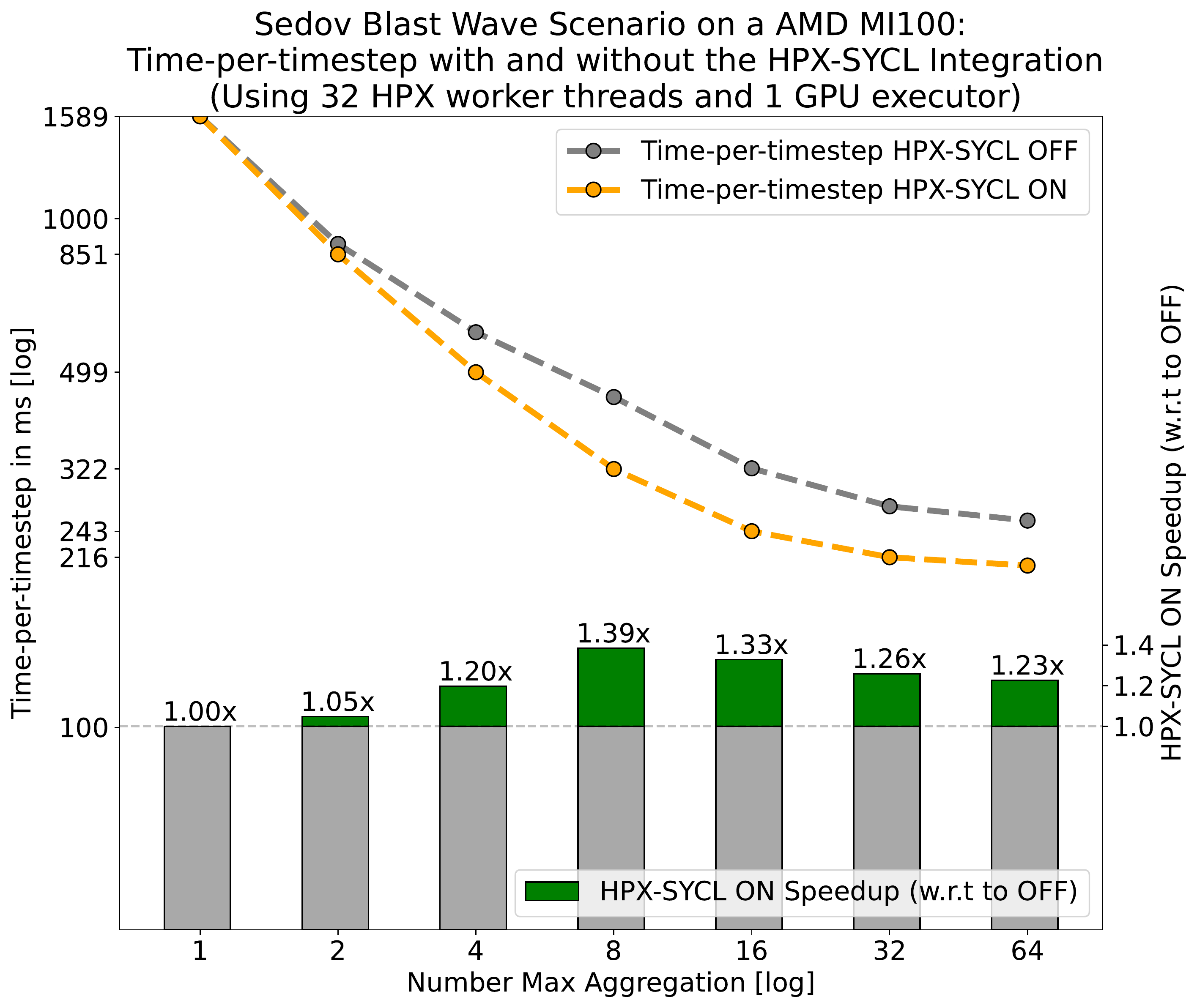}  
}
\hspace*{-0.1cm} 
\subfloat[\label{fig:integration-event-polling-mi100-3}MI100: Best combinations] {
\centering
  \includegraphics[width=.33\textwidth]{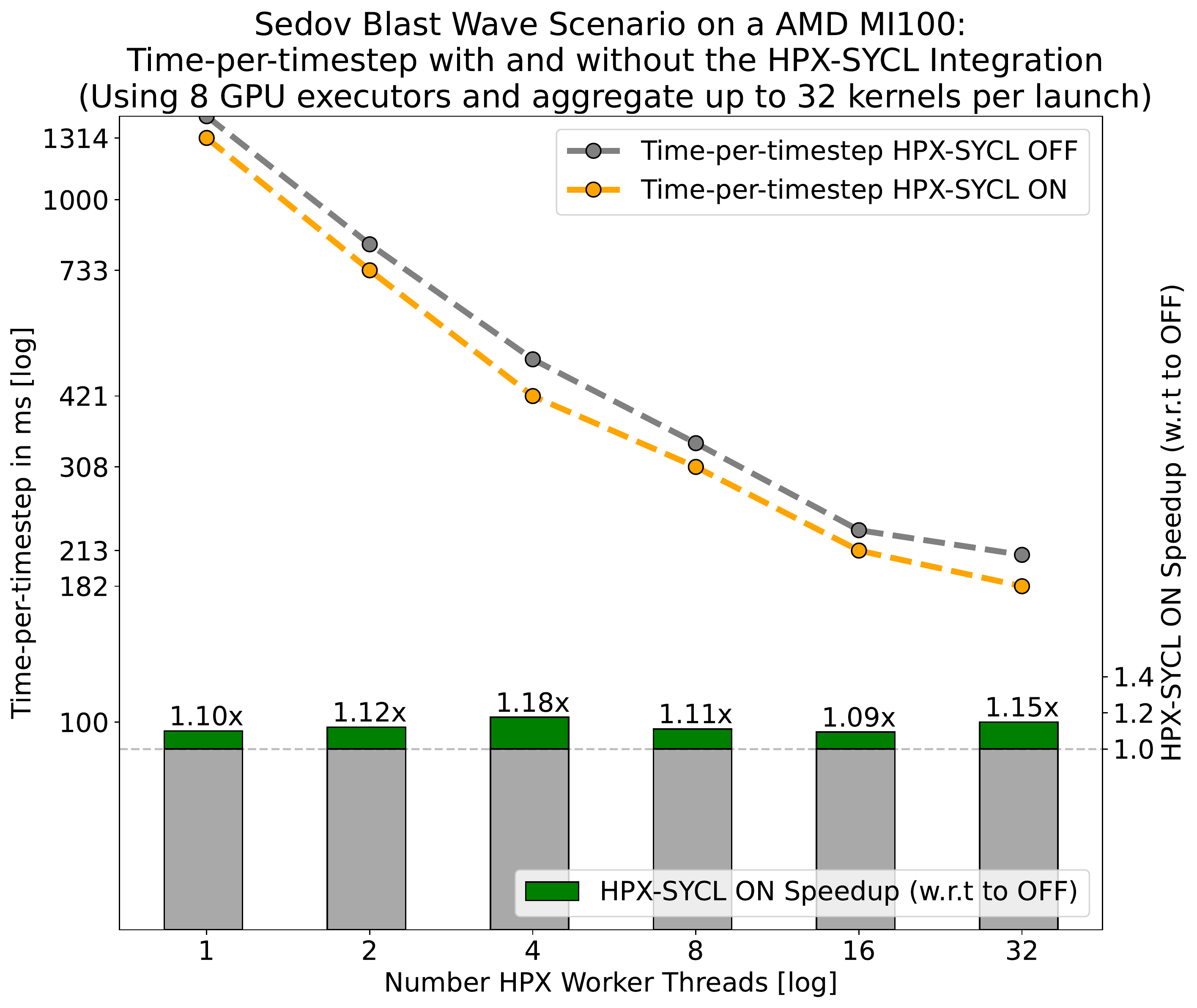}  
}
\caption{Runs with the SYCL-Integration (event polling version) turned on (orange) and turned off (gray). In \ref{fig:integration-event-polling-a100-1} and \ref{fig:integration-event-polling-mi100-1} we do not use the dynamic work aggregation and instead only increase the number of GPU executors. In \ref{fig:integration-event-polling-a100-2} and \ref{fig:integration-event-polling-mi100-2} we only use one executor, but increase the maximum number of kernels aggregated into one kernel launch. In \ref{fig:integration-event-polling-a100-3} and \ref{fig:integration-event-polling-mi100-3} we see the best combinations. The Kokkos optimization patch is applied to all runs.}
\label{fig:integration-event-polling}
\end{figure*}
\subsubsection{Test 2 - Performance Impact of the Event Polling HPX-SYCL Integration}

\begin{figure*}[t]
\centering
\subfloat[\label{fig:kokkos-patch-a100-1}A100: Increasing Number of GPU executors] {
\centering
  \includegraphics[width=.33\textwidth]{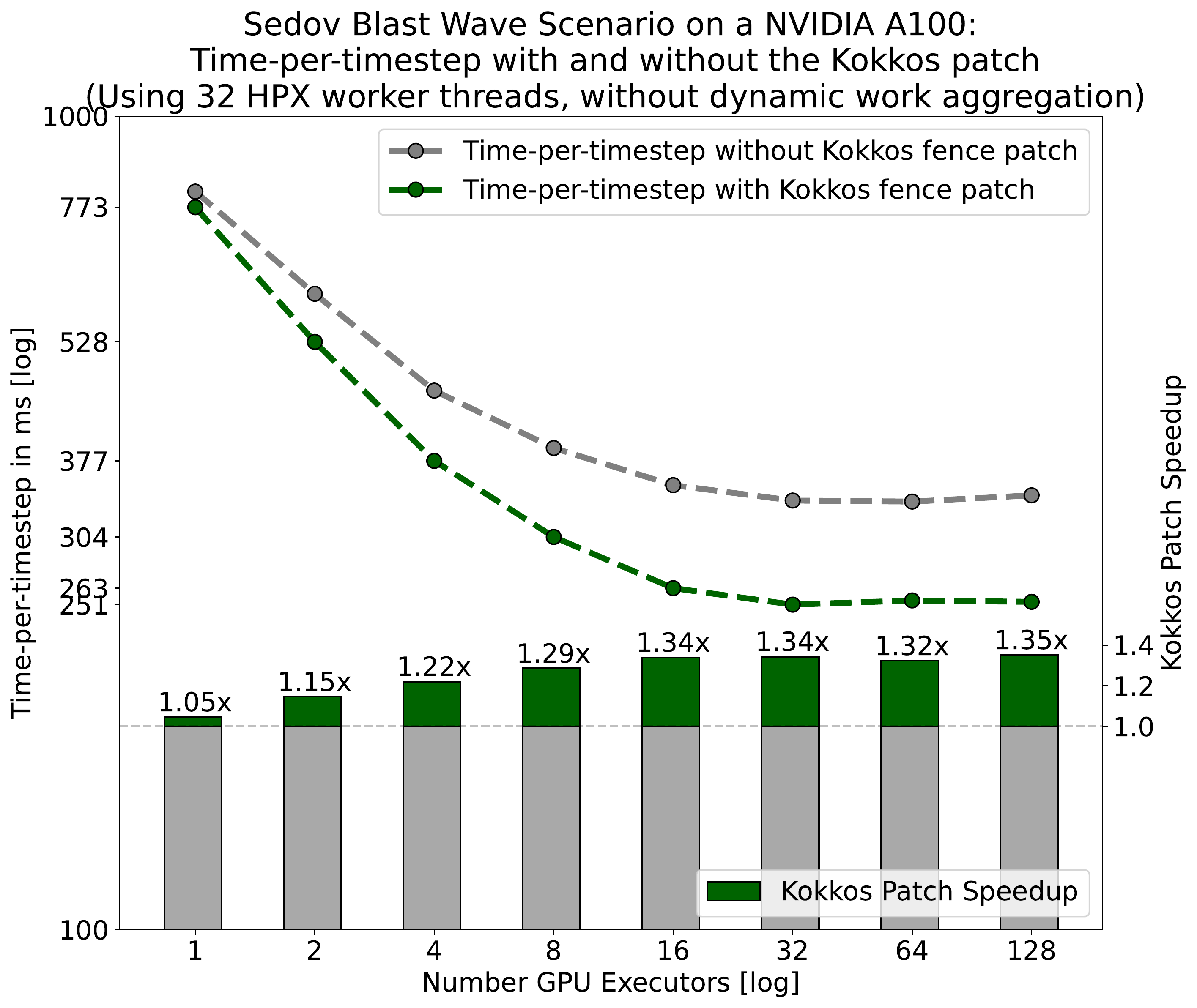}  
}
\hspace*{-0.1cm} 
\subfloat[\label{fig:kokkos-patch-a100-2}A100: Increasing number of kernels aggregated] {
\centering
  \includegraphics[width=.33\textwidth]{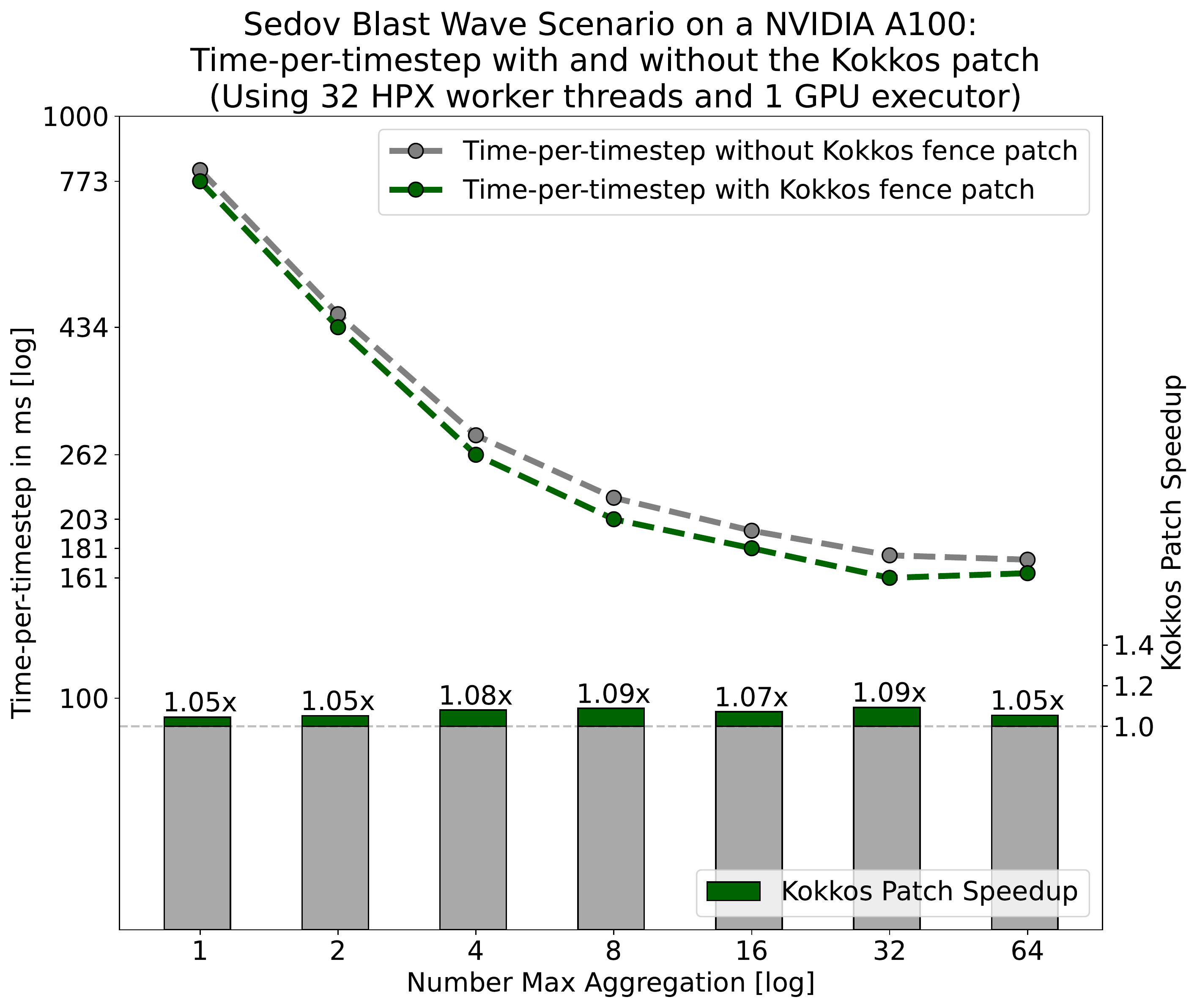}  
}
\hspace*{-0.1cm} 
\subfloat[\label{fig:kokkos-patch-a100-3}A100: Best combinations] {
\centering
  \includegraphics[width=.33\textwidth]{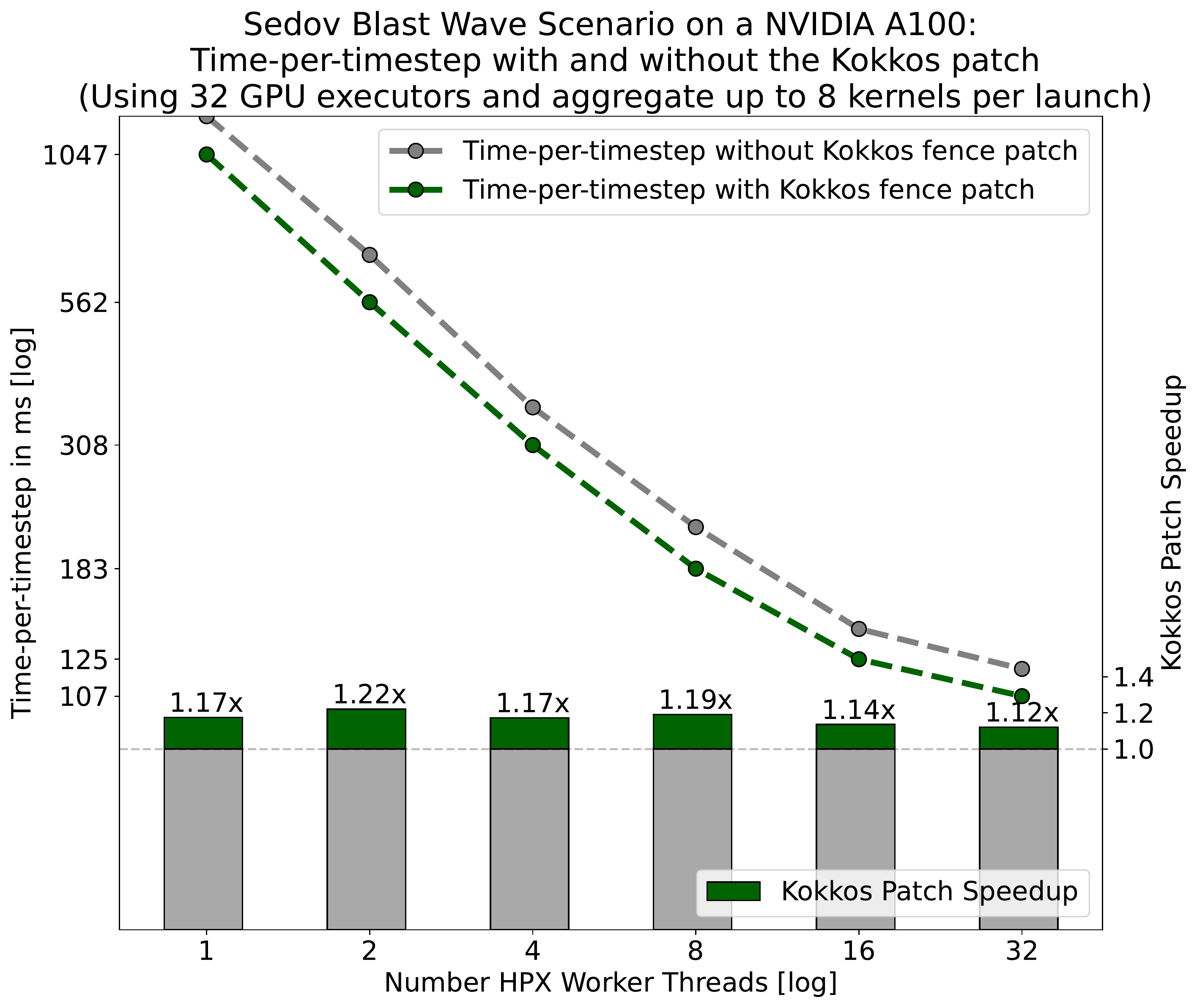}  
}
\hfill
\centering
\subfloat[\label{fig:kokkos-patch-mi100-1}MI100: Increasing Number of GPU executors] {
\centering
  \includegraphics[width=.33\textwidth]{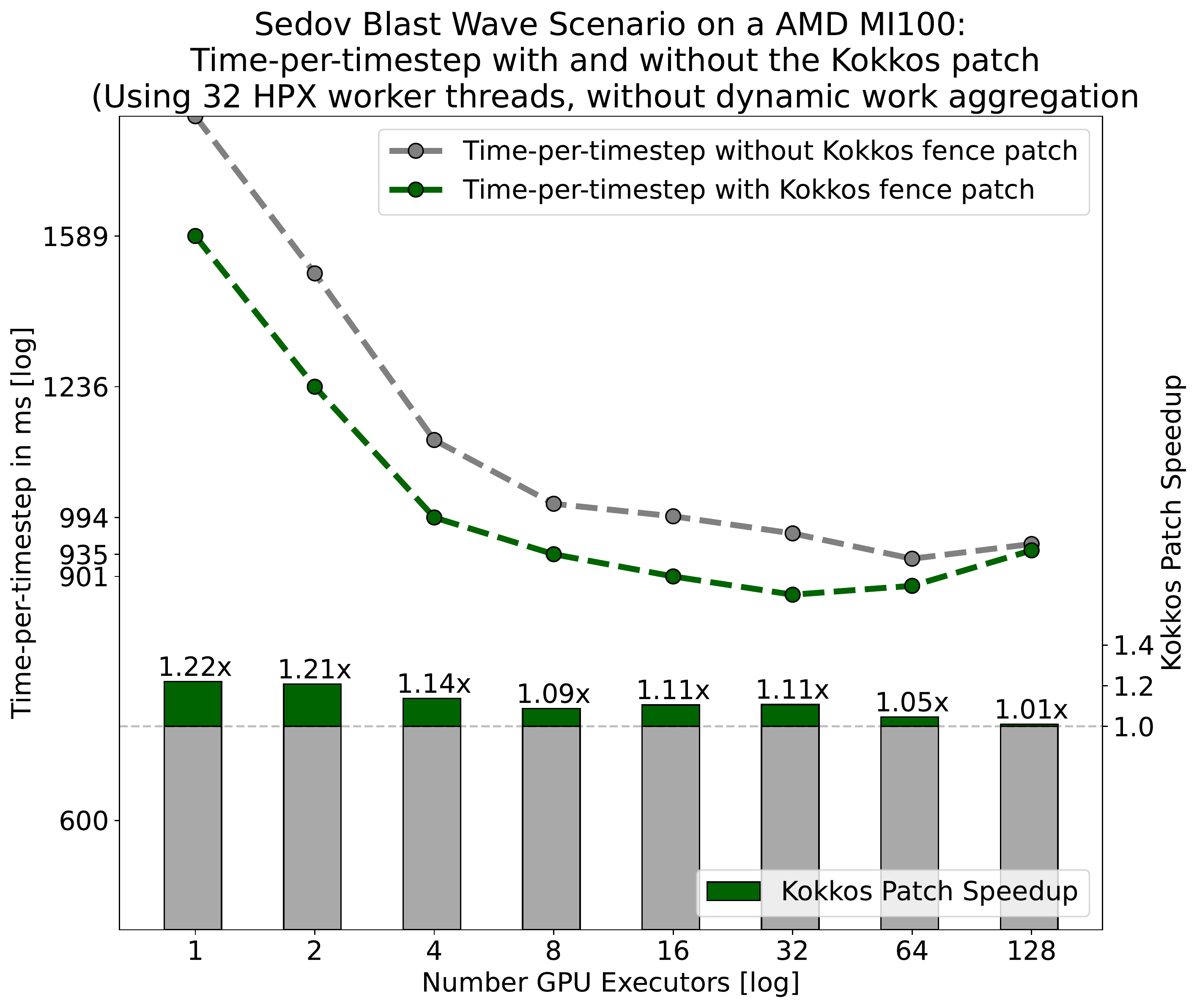}  
}
\hspace*{-0.1cm} 
\subfloat[\label{fig:kokkos-patch-mi100-2}MI100: Increasing number of kernels aggregated] {
\centering
  \includegraphics[width=.33\textwidth]{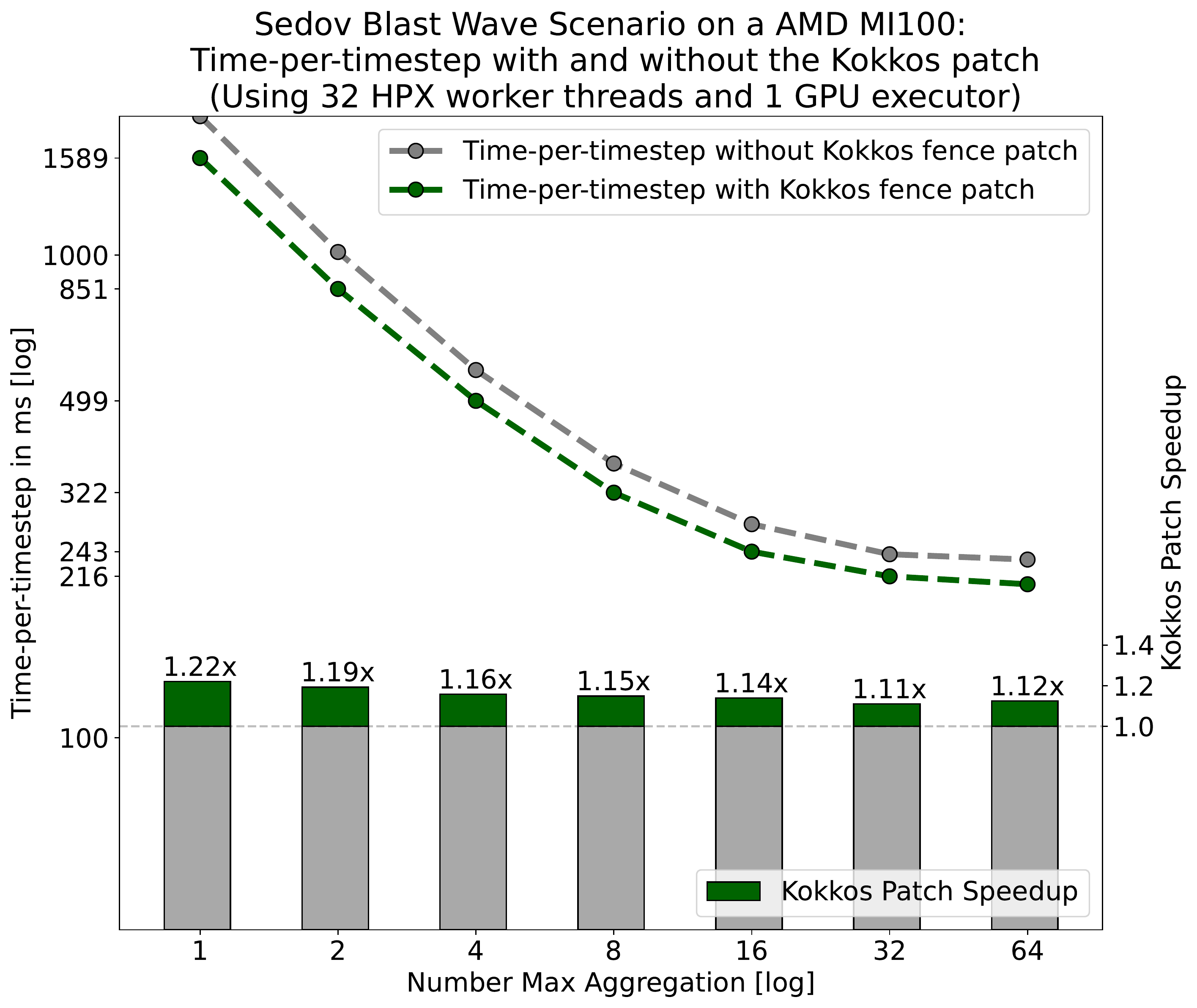}  
}
\hspace*{-0.1cm} 
\subfloat[\label{fig:kokkos-patch-mi100-3}MI100: Best combinations] {
\centering
  \includegraphics[width=.33\textwidth]{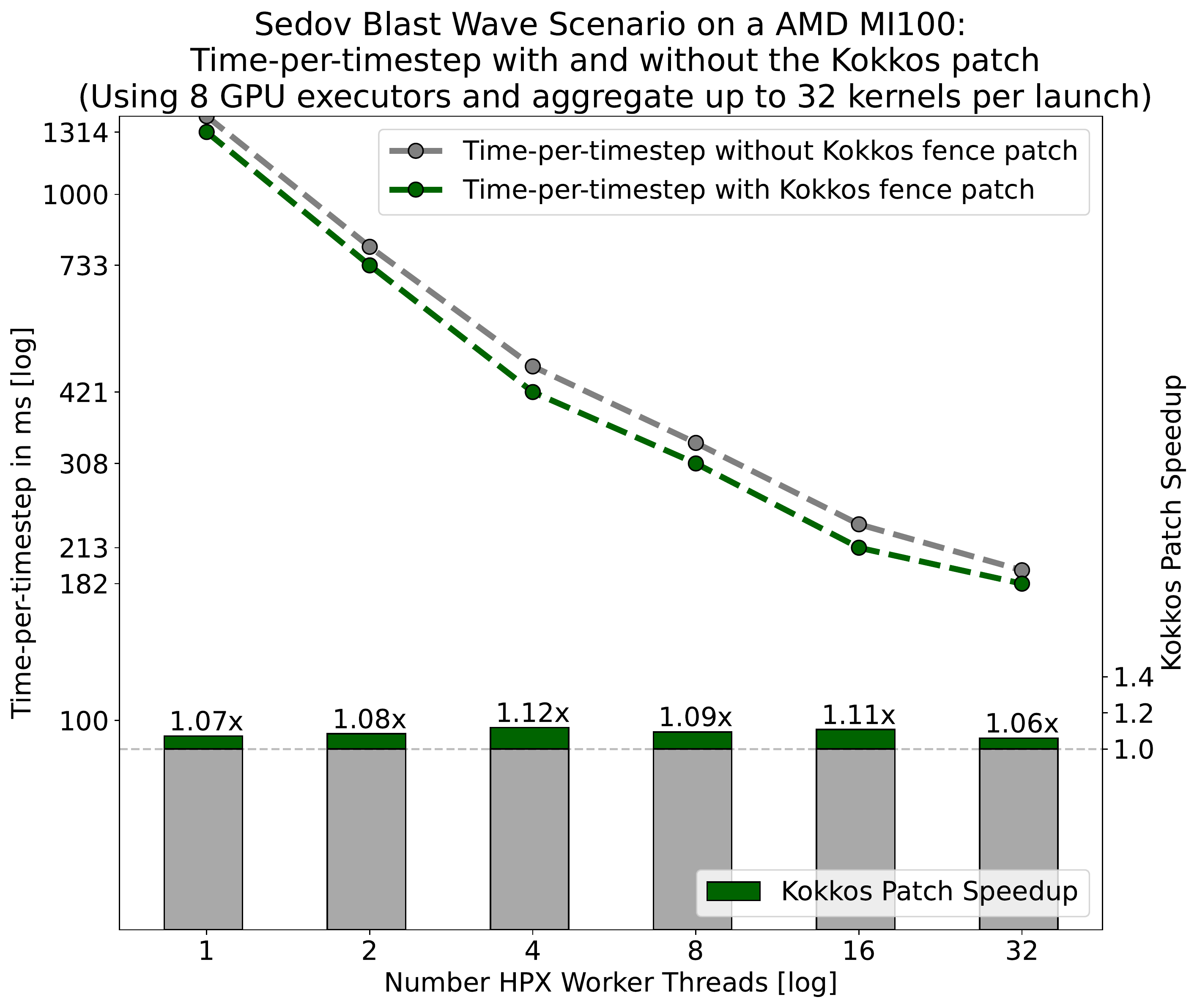}  
}

\caption{Runs with the applied Kokkos optimization patch (green) and without the patch (gray). In \ref{fig:kokkos-patch-a100-1} and \ref{fig:kokkos-patch-mi100-1} we do not use the dynamic work aggregation and instead only increase the number GPU executors. In \ref{fig:kokkos-patch-a100-2} and \ref{fig:kokkos-patch-mi100-2} we only use one executor but increase the maximum number of kernels aggregated into one kernel launch. In \ref{fig:kokkos-patch-a100-3} and \ref{fig:kokkos-patch-mi100-3}, we see the best combinations. The HPX-SYCL integration (event polling version) is on for all runs.}
\label{fig:kokkos-patch-a100}
\end{figure*}
Given the disappointing results of the \lstinline{host_task} integration, it is clear that we
need an alternative. The event polling version we implemented does not suffer
from the same drawbacks, as everything is handled by the HPX threads themselves
in a way that is optimized for multi-threaded usage. However, the continuous
polling is adding a different kind of overhead, making it important to also
check this integration by running Octo-Tiger once with it enabled and once
while it is disabled.

Hence, we are now repeating the same tests as in the previous section, but
using the event polling version of our integration. The results can be found in
Figure~\ref{fig:integration-event-polling}. This time, we can achieve clear
speedups, especially for the graphs that only use one GPU executor (\ref{fig:integration-event-polling-a100-2}, \ref{fig:integration-event-polling-mi100-2}) but
with the work aggregation enabled. Here, we benefit from the aggregation the most,
as CPU-time becomes increasingly more valuable as the worker threads are busy
coordinating the work aggregation on top of their other tasks. Even for the
graphs with the best combinations (\ref{fig:integration-event-polling-a100-3}, \ref{fig:integration-event-polling-mi100-3}) we see clear benefits. Slightly more
so when using fewer HPX worker threads, however, even when using all $32$ workers with the best
combination, we see a speedup of $1.11x$ on the A100 node and one of $1.15x$ on
the MI100 node. Note that the best combination on the MI100 node is different
from the A100 one, as we benefit less from concurrent GPU executors on the AMD
GPU, and instead just use $8$ GPU executors with up to $32$ kernels aggregated.
We have seen a similar effect in previous tests using HIP on this machine,
which was one of the original triggers for us to implement the dynamic work
aggregation executor~\cite{daiss2022aggregation}.

\subsubsection{Test 3 - Performance Impact of the Kokkos Modifications}

The last two tests were about the HPX-SYCL integration, but always had the
Kokkos optimization patch applied (both the runs with and without integration).
In this test, we always have the (event polling) HPX-SYCL integration enabled,
but toggle whether we use the Kokkos modification patch. As mentioned, this
patch basically just skips a few barriers in case the Kokkos SYCL execution
space is using an \lstinline{in_order} queue. When using it with Octo-Tiger we did not
notice any deviation in the actual results of our tests. However, it made a
performance impact.
The results can be found in Figure~\ref{fig:kokkos-patch-a100}. The patch is
consistently beneficial, with us reaching speedups of around $1.06$ and $1.12$
for the best combination runs. The speedup on the A100 node without dynamic
work aggregation (\ref{fig:kokkos-patch-a100-1}) stands out. We benefit more from the concurrent GPU
executors on NVIDIA hardware, so keeping them
better fed with kernels (without any blocking) yields a larger advantage.
Overall, applying the patch is beneficial in all tested configurations.

\subsubsection{Test 4 - Performance Using Different Execution Backends}
\begin{figure}[t]
\subfloat[\label{fig:a100-backend}Best runs on the NVIDIA A100] {
\centering
  \includegraphics[width=.23\textwidth]{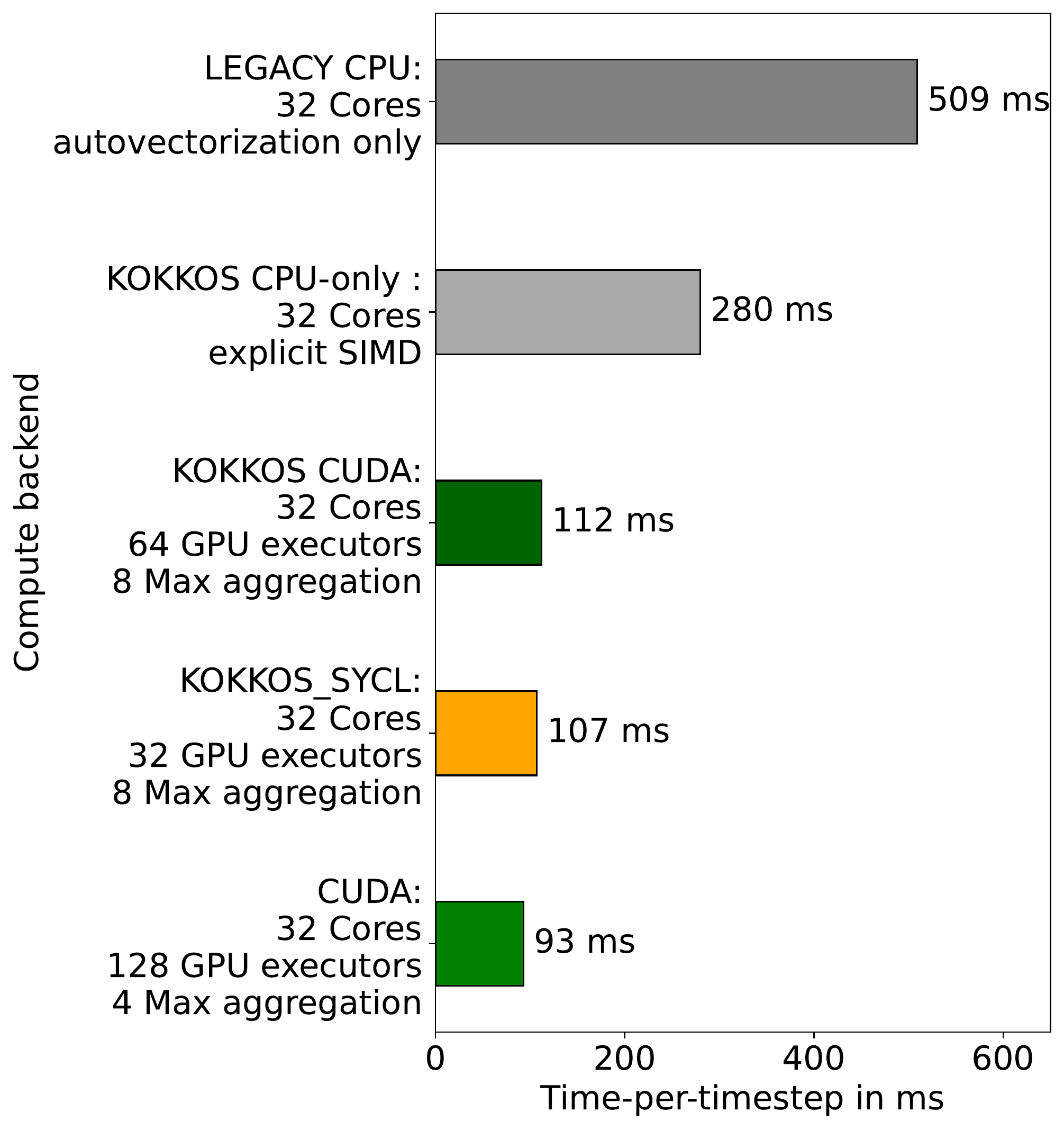}  
}
\hspace*{-0.1cm} 
\subfloat[\label{fig:mi100-backend}Best runs on the AMD MI100] {
\centering
  \includegraphics[width=.23\textwidth]{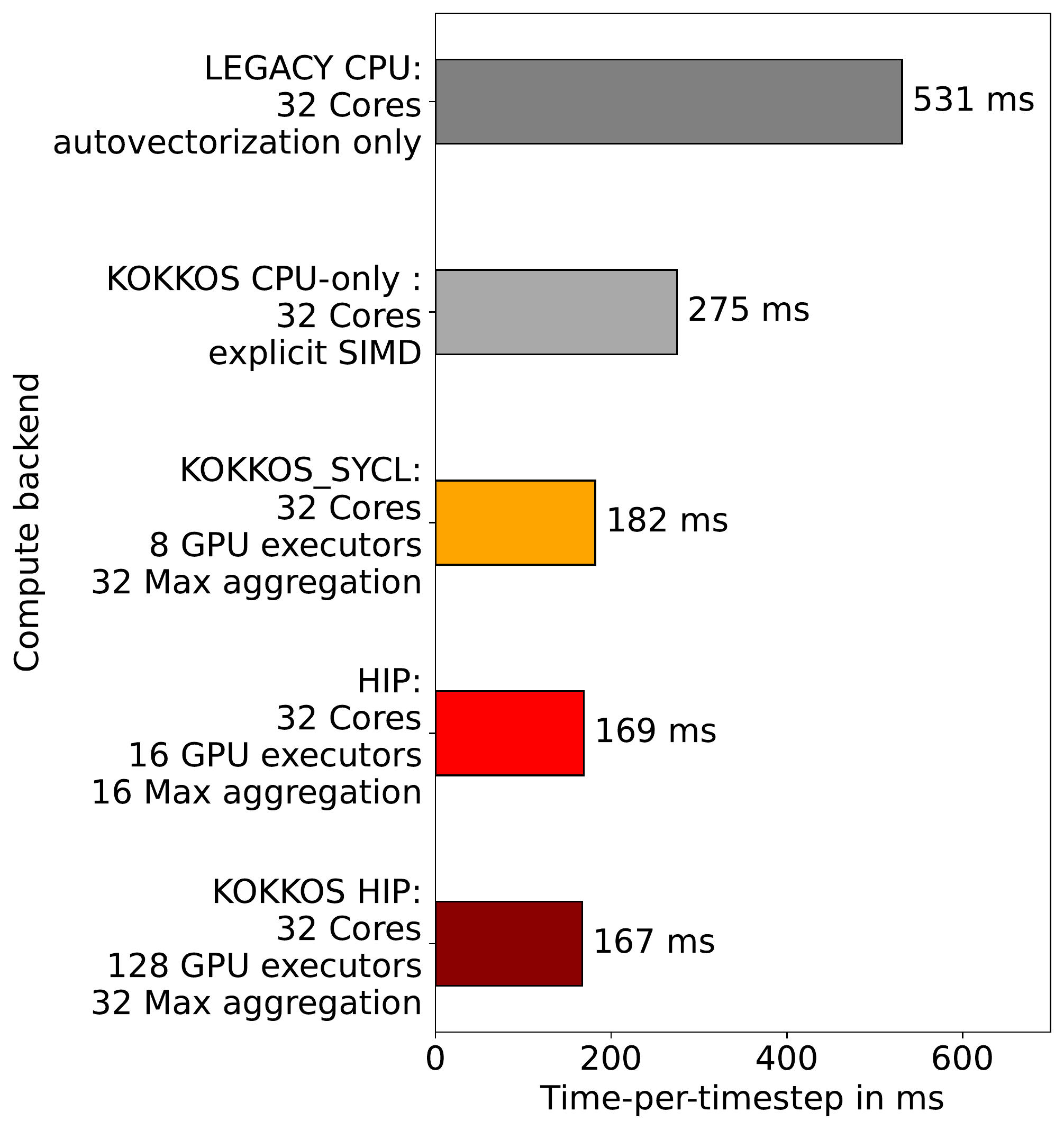}  
}
\caption{Sedov Blast Scenario with Octo-Tiger using all the available backends on both the A100 node (\ref{fig:a100-backend}) and the MI100 node (\ref{fig:mi100-backend}). We used the best combinations for each backend and added CPU-only runs for comparison.}
\label{fig:backend-overview}
\end{figure}
On the A100 node we can now use multiple backends for the kernels: plain CUDA,
Kokkos using the CUDA execution space, and Kokkos using the SYCL execution
space. This warrants a closer look at how these different backends perform
with Octo-Tiger using the same scenario and the same hardware.
The results for this can be found in Figure~\ref{fig:backend-overview} (both
for the A100 and MI100 node). The (event polling) HPX-SYCL integration is ON and the Kokkos
optimization patch is applied. Interestingly, in this scenario, the Kokkos
SYCL backend seems to be competitive compared to its CUDA equivalent. From
what we are able to tell from the profiler output, the average runtime per
kernel is actually better with SYCL. However, we are losing some runtime again
in the overhead, since the HPX-CUDA integration is using an event pool, while
we have to get our SYCL events from the SYCL runtime, resulting in frequent
creation and destruction of these SYCL events. Notably, the Kokkos CUDA
execution space version here is a bit slower than it was in previous work,
where we used Clang 12 instead of DPC\texttt{++}~\cite{daiss2022aggregation}
and an older version of Octo-Tiger which did not yet include the SIMD types
within the Kokkos kernels (which might add a bit of overhead to the GPU
execution).
It is noteworthy, that independent of the GPU backend used, Octo-Tiger performs
better on the A100 GPU than it does on the MI100 one. This shows that the
kernels themselves need more optimization for the AMD GPU.


\section{Related work}
\label{sec:related:work}
From the application perspective, other astrophysics codes support SYCL's abstraction layer as well. \textit{DPEcho}\footnote{\url{https://github.com/LRZ-BADW/DPEcho}}, a code for general relativistic magneto hydrodynamics, uses Intel MPI and hipSYCL. \textit{ARGOT}, a radiative transfer code, uses Intel DPC\texttt{++} to support GPUs (Intel CPU + NVIDIA GPU) and Intel FPGAs~\cite{kashino2022multi}. However, the applications, \emph{e.g.}\ black holes and cosmology, of these two codes are different from Octo-Tiger.

From the integration of SYCL within asynchronous multitask systems, we focus on the integration of the asynchronous tasks of the SYCL  API. One could just call the SYCL  API on one thread, block the thread while waiting for the event. However, we are interested in integrating the SYCL  API call into the asynchronous execution graph. Chapel~\cite{chamberlain2007parallel} supports Intel DPC\texttt{++} as one of their GPU API modules since October 21. Unitah~\cite{germain2000uintah} support SYCL via Kokkos SYCL \cite{10.1145/3539781.3539794}. Chiu et al. compared the SYCL default task graph with the CUDA graph execution model\footnote{\url{https://docs.nvidia.com/cuda/cuda-c-programming-guide/index.html\#cuda-graphs}} for large-scale machine learning work loads~\cite{chiu2022experimental}.  Other notable AMTs are: Charm\texttt{++}~\cite{kale1993charm++}, Legion~\cite{bauer2012legion}, and PaRSEC~\cite{bosilca2013parsec}. A detailed comparison is given in~\cite{thoman2018taxonomy}. 
Focusing on the programming model, Charm\texttt{++} and HPX are very close; however, HPX conforms to the C\texttt{++} standard and Charm\texttt{++} is a library that uses the C\texttt{++} programming language. The overheads using HPX and Charm\texttt{++} are compared with MPI and OpenMP in~\cite{https://doi.org/10.48550/arxiv.2207.12127}.

\section{Conclusion and Future Work}
\label{sec:conclusion}
For this work, we began making Octo-Tiger and HPX itself compatible with SYCL.
To do so, we introduced multiple software changes, most important of which is
the HPX-SYCL integration. 

Integrating these two frameworks certainly seems like an odd choice initially,
as both contain similar functionality for expressing execution graphs (one
using SYCL events, the other using HPX futures). However, there are
advantages, as both frameworks have their own specialties: With SYCL we can
quickly build the asynchronous dependencies between GPU kernels and
data-transfers, as shown in Octo-Tiger's hydro solver, where we first schedule
5 GPU kernels and their associated CPU-GPU data transfers for each sub-grid,
before HPX gets involved again. HPX, on the other hand, excels in scheduling
asynchronous CPU work, whether it is on the local compute-node or a remote one.
For instance, after scheduling the GPU work with SYCL, we use HPX by getting a
future when the GPU work for a sub-grid is done, then use this future to
schedule the CPU post-processing and the communication with neighboring
sub-grids without ever blocking the CPU host threads. Although we only looked
at single-node runs in this work, this should become even more important in
distributed runs, as the interleaving of computation and communication becomes
increasingly more important the more nodes we use, especially in tree-based
codes like Octo-Tiger.

While we implemented the HPX-SYCL integration using both the \lstinline{host_task} SYCL feature and event polling, in the end, the version
using event polling within the HPX scheduler proved to be clearly superior in
our tests. Overall, the speedups in Octo-Tiger when using the integration are
modest, yet noticeable, ranging from $1.11x$ to $1.15x$ for the best
configurations (using $32$ HPX worker threads and the best combination of GPU executors and
maximum number of aggregated kernels). 
Furthermore, the Kokkos optimization
patch we used shows promise, though at the point of submission, we think it
requires further testing with codes beyond Octo-Tiger before upstreaming it.
Between these Kokkos changes and the SYCL integration, Octo-Tiger's overall
runtime behavior when using the Kokkos SYCL execution space becomes comparable to the
one using the Kokkos CUDA execution space.

For future work, we would like to make use of two opportunities that the
groundwork in this paper enables: First, since we updated and optimized
Octo-Tiger's toolchain to work with SYCL, we can now realistically target the
upcoming Intel GPUs for future Octo-Tiger runs. Hence, we would like to
re-run the tests shown here on the respective Intel hardware in the future.
Secondly, one of the major advantages of using HPX is that it enables
distributed runs. We would like to revisit the benefit of the HPX-SYCL
integration in more complex, distributed scenarios on Perlmutter, as blocking
the cores by waiting for SYCL results (without the integration) should have a
more profound negative performance impact here as it can impact the
interleaving of computation and communication.


\bibliographystyle{plain}
\bibliography{main}

\appendix

\section{Supplementary materials}
The Octo-Tiger build scripts are available on GitHub\footnote{\url{https://github.com/STEllAR-GROUP/OctoTigerBuildChain}}. For the runs on the A100 node we used the git branch \textit{sycl\_toolchain} of this repository, for the ones on the MI100 node \textit{sycl\_toolchain\_hip}.

\end{document}